\documentclass[twocolumn]{aastex63}

\usepackage{amssymb,amsmath,amsthm}
\usepackage{graphicx}
\usepackage{mathrsfs}
\usepackage{booktabs}
\usepackage{multirow}
\usepackage{empheq}
\usepackage{verbatim}
\usepackage{mathtools}
\usepackage{hyperref}
\usepackage{xcolor}
\usepackage{xspace}
\usepackage{comment}
\usepackage{dirtytalk}

\def\m87{M87$^*$\xspace}
\def\sgra{Sgr~A$^*$\xspace}

\newcommand{\uv}{$(u,v)$\xspace}
\newcommand{\ff}{$\mu_{\rm ff}$\xspace}
\newcommand{\uas}{$\mu$as\xspace}

\newcommand{\phaseone}{Phase\,1\xspace}
\newcommand{\phasetwo}{Phase\,2\xspace}

\def\uas{\ensuremath{\mu}as\xspace} 

\defcitealias{Narayan_1995b}{NY95}
\defcitealias{Mahadevan_1997}{M97}
\defcitealias{Shankar_2009}{SWM}

\defcitealias{Paper1}{EHTC I}
\defcitealias{Paper2}{EHTC II}
\defcitealias{Paper3}{EHTC III}
\defcitealias{Paper4}{EHTC IV}
\defcitealias{Paper5}{EHTC V}
\defcitealias{Paper6}{EHTC VI}

\begin{document}

\nocite{Paper1}
\nocite{Paper2}
\nocite{Paper3}
\nocite{Paper4}
\nocite{Paper5}
\nocite{Paper6}

\title{Reference Array and Design Consideration for the next-generation Event Horizon Telescope}

\correspondingauthor{Sheperd~S.~Doeleman}
\email{sdoeleman@cfa.harvard.edu}

\author[0000-0002-9031-0904]{Sheperd~S.~Doeleman}
\affiliation{Center for Astrophysics $|$ Harvard \& Smithsonian, 60 Garden Street, Cambridge, MA 02138, USA}
\affiliation{Black Hole Initiative at Harvard University, 20 Garden Street, Cambridge, MA 02138, USA}

\author[0000-0002-9290-0764]{John Barrett}
\affiliation{Massachusetts Institute of Technology Haystack Observatory, 99 Millstone Road, Westford, MA 01886, USA}

\author[0000-0002-9030-642X]{Lindy~Blackburn}
\affiliation{Center for Astrophysics $|$ Harvard \& Smithsonian, 60 Garden Street, Cambridge, MA 02138, USA}
\affiliation{Black Hole Initiative at Harvard University, 20 Garden Street, Cambridge, MA 02138, USA}

\author{Katherine Bouman}
\affiliation{California Institute of Technology, 1200 East California Boulevard, Pasadena, CA 91125 USA}

\author[0000-0002-3351-760X]{Avery E. Broderick}
\affiliation{Perimeter Institute for Theoretical Physics, 31 Caroline Street North, Waterloo, ON, N2L 2Y5, Canada}
\affiliation{Department of Physics and Astronomy, University of Waterloo, 200 University Avenue West, Waterloo, ON, N2L 3G1, Canada}
\affiliation{Waterloo Centre for Astrophysics, University of Waterloo, Waterloo, ON, N2L 3G1, Canada}

\author{Ryan Chaves}
\affiliation{Center for Astrophysics $|$ Harvard \& Smithsonian, 60 Garden Street, Cambridge, MA 02138, USA}

\author[0000-0002-7128-9345]{Vincent L. Fish}
\affiliation{Massachusetts Institute of Technology Haystack Observatory, 99 Millstone Road, Westford, MA 01886, USA}

\author{Garret Fitzpatrick}
\affiliation{Center for Astrophysics $|$ Harvard \& Smithsonian, 60 Garden Street, Cambridge, MA 02138, USA}

\author{Antonio Fuentes}
\affiliation{Instituto de Astrof\'{\i}sica de Andaluc\'{\i}a-CSIC, Glorieta de la Astronom\'{\i}a s/n, E-18008 Granada, Spain}

\author{Mark Freeman}
\affiliation{Center for Astrophysics $|$ Harvard \& Smithsonian, 60 Garden Street, Cambridge, MA 02138, USA}

\author[0000-0003-4190-7613]{Jos\'e L. G\'omez}
\affiliation{Instituto de Astrof\'{\i}sica de Andaluc\'{\i}a-CSIC, Glorieta de la Astronom\'{\i}a s/n, E-18008 Granada, Spain}

\author{Kari Haworth}
\affiliation{Center for Astrophysics $|$ Harvard \& Smithsonian, 60 Garden Street, Cambridge, MA 02138, USA}

\author{Janice Houston}
\affiliation{Center for Astrophysics $|$ Harvard \& Smithsonian, 60 Garden Street, Cambridge, MA 02138, USA}

\author{Sara Issaoun}
\affiliation{Center for Astrophysics $|$ Harvard \& Smithsonian, 60 Garden Street, Cambridge, MA 02138, USA}

\author[0000-0002-4120-3029]{Michael D. Johnson}
\affiliation{Center for Astrophysics $|$ Harvard \& Smithsonian, 60 Garden Street, Cambridge, MA 02138, USA}
\affiliation{Black Hole Initiative at Harvard University, 20 Garden Street, Cambridge, MA 02138, USA}

\author[0000-0002-6156-5617]{Mark Kettenis}
\affiliation{Joint Institute for VLBI ERIC (JIVE), Oude Hoogeveensedijk 4, 7991 PD Dwingeloo, The Netherlands}

\author[0000-0002-5635-3345]{Laurent Loinard}
\affiliation{Instituto de Radioastronomía y Astrofísica, Universidad Nacional Autónoma de México, Morelia 58089, México}
\affiliation{Instituto de Astronomía, Universidad Nacional Autónoma de México (UNAM), Apdo Postal 70-264, Ciudad de México, México}

\author[0000-0001-6920-662X]{Neil Nagar}
\affiliation{Astronomy Department, Universidad de Concepción, Casilla 160-C, Concepción, Chile}

\author[0000-0002-4723-6569]{Gopal Narayanan}
\affiliation{Department of Astronomy, University of Massachusetts, 01003, Amherst, MA, USA}

\author{Aaron Oppenheimer}
\affiliation{Center for Astrophysics $|$ Harvard \& Smithsonian, 60 Garden Street, Cambridge, MA 02138, USA}

\author[0000-0002-7179-3816]{Daniel~C.~M.~Palumbo}
\affiliation{Center for Astrophysics $|$ Harvard \& Smithsonian, 60 Garden Street, Cambridge, MA 02138, USA}
\affiliation{Black Hole Initiative at Harvard University, 20 Garden Street, Cambridge, MA 02138, USA}

\author[0000-0002-6021-9421]{Nimesh Patel}
\affiliation{Center for Astrophysics $|$ Harvard \& Smithsonian, 60 Garden Street, Cambridge, MA 02138, USA}

\author[0000-0002-5278-9221]{Dominic W. Pesce}
\affiliation{Center for Astrophysics $|$ Harvard \& Smithsonian, 60 Garden Street, Cambridge, MA 02138, USA}
\affiliation{Black Hole Initiative at Harvard University, 20 Garden Street, Cambridge, MA 02138, USA}

\author[0000-0002-5779-4767]{Alexander W. Raymond}
\affiliation{Black Hole Initiative at Harvard University, 20 Garden Street, Cambridge, MA 02138, USA}
\affiliation{Center for Astrophysics $|$ Harvard \& Smithsonian, 60 Garden Street, Cambridge, MA 02138, USA}

\author[0000-0001-5461-3687]{Freek Roelofs}
\affiliation{Center for Astrophysics $|$ Harvard \& Smithsonian, 60 Garden Street, Cambridge, MA 02138, USA}
\affiliation{Black Hole Initiative at Harvard University, 20 Garden Street, Cambridge, MA 02138, USA}

\author{Ranjani Srinivasan}
\affiliation{Center for Astrophysics $|$ Harvard \& Smithsonian, 60 Garden Street, Cambridge, MA 02138, USA}

\author[0000-0003-3826-5648]{Paul Tiede}
\affiliation{Center for Astrophysics $|$ Harvard \& Smithsonian, 60 Garden Street, Cambridge, MA 02138, USA}
\affiliation{Black Hole Initiative at Harvard University, 20 Garden Street, Cambridge, MA 02138, USA}

\author[0000-0002-4603-5204]{Jonathan Weintroub}
\affiliation{Center for Astrophysics $|$ Harvard \& Smithsonian, 60 Garden Street, Cambridge, MA 02138, USA}
\affiliation{Black Hole Initiative at Harvard University, 20 Garden Street, Cambridge, MA 02138, USA}

\author[0000-0002-8635-4242]{Maciek Wielgus}
\affiliation{Max-Planck-Institut f\"ur Radioastronomie, Auf dem H\"ugel 69, D-53121 Bonn, Germany}

\begin{abstract}
We describe the process to design, architect, and implement a transformative enhancement of the Event Horizon Telescope (EHT).  This program - the next-generation Event Horizon Telescope (ngEHT) - will form a networked global array of radio dishes capable of making high-fidelity real-time movies of supermassive black holes (SMBH) and their emanating jets.  This builds upon the EHT principally by deploying additional modest-diameter dishes to optimized geographic locations to enhance the current global mm/submm wavelength Very Long Baseline Interferometric (VLBI) array, which has, to date, utilized mostly pre-existing radio telescopes.  The ngEHT program further focuses on observing at three frequencies simultaneously for increased sensitivity and Fourier spatial frequency coverage.  Here, the concept, science goals, design considerations, station siting and instrument prototyping is discussed, and a preliminary reference array to be implemented in phases is described.
\end{abstract}

\section{Introduction}

On April 10, 2019, the Event Horizon Telescope project (EHT) released images of the supermassive black hole at the heart of galaxy M87 \citep{Paper1, Paper2, Paper3, Paper4, Paper5, Paper6}.  The observed ring of emission, formed by radio waves lensed in the gravitational field of a 6.5 billion solar mass black hole, has dimensions that match the predictions of General Relativity.  Images of \sgra, the 4 million solar mass black hole at the center of the Milky Way, also exhibit a ring morphology with diameters anticipated by theory \citep{sgr_paper1, sgr_paper2, sgr_paper3, sgr_paper4, sgr_paper5, sgr_paper6}.  These results confirm that the EHT has observed the strong gravitational lensing signature of supermassive black holes \citep{Bardeen_1973,Luminet_1979, Falcke_2000,Takahashi_2004,Broderick_2006}, and these images have opened a new field of precision black hole studies on horizon scales.

This work built upon decades of technical development and precursor observations.  Pioneering first Very Long Baseline Interferometry (VLBI) experiments at wavelengths of 1.3mm \citep{Padin1990, Krichbaum1998} demonstrated that observations with the required resolution were possible at frequencies where AGN are likely to be optically thin.  Discovery of horizon-scale structure in both Sgr~A* and M87 with purpose-built ultra-high bandwidth systems on early EHT arrays \citep{Doeleman_2008, Doeleman_2012} confirmed that imaging these sources was feasible.  Subsequent observations revealed time-variability and ordered magnetic fields on Schwarzschild radius dimensions \citep{Fish2011, Johnson2015}.  Emergence of the EHT to a full imaging array grew from building community support through a decadal review processes \citep{Doeleman_2009}, efforts to modify large scale international facilities, such as ALMA, through global cooperation \citep{Doeleman_2010,Matthews_2018}, and work to enable VLBI capability at the most remote observatories on the planet \citep{Inoue2014_glt, Kim2018_spt}. Over the course of two decades, all the technical, logistical, organizational and analytical aspects of the full EHT were implemented by an expert team that grew from a few 10’s to over 200 collaborators worldwide.  

Building upon this legacy, the next-generation EHT (ngEHT) provides a roadmap to greatly accelerate the development of the EHT, envisaging a transformative new instrument capable of delivering real-time black hole movies.  Where the EHT used existing mm/sub-mm facilities to form a first imaging array, the ngEHT will take the next step by designing and locating new dishes to optimize performance and scientific return.  This vision offers excellent opportunities to engage the curious public on many levels.  It is estimated that over a billion people have now seen the M87 image \citep{Christensen2019}.  We anticipate that the long term public and STEM education engagement as the ngEHT builds to its goal of black hole ‘cinema’ will be similar in scope. 

For the purposes of this paper, the term \say{ngEHT} is used to describe a program to explore and define a long-term plan to enhance the EHT to realize a new set of transformative science goals.  This paper describes that vision by outlining improvements in bandwidth, frequency range, new antenna deployment and new operating modes that enable increases in angular resolution, Fourier spatial frequency coverage, sensitivity and temporal resolution.  For brevity, \say{ngEHT} will also be used as shorthand for the future arrays that will emerge through these plans, as well as for the constellation of improvements that constitute the ngEHT concept.  

Technical advances in several areas make design and implementation of the ngEHT within this decade a realistic goal.  Over most of the past two decades, the bandwidth of VLBI systems has kept pace with Moore's Law - a doubling of capacity and speed approximately every 18 month (see, e.g.,\,\citet{Paper2} and \autoref{fig:eht_bandwidth} below).  This is primarily due to the migration of VLBI instrumentation development to designs that adopt industry-standard components, including CPUs, Analog to Digital Converters, Field Programmable Gate Arrays (FPGA) and commercial data transmission protocols (e.g., \citet{Vertatschitsch2015}).   The increased bandwidth of these components and systems match the analog bandwidth improvements planned for international and national submm facilities, including ALMA \citep{ALMA2030} and the Submillimeter Array \citep{wSMA_2020}.  Meanwhile, the transport of larger data volumes captured by next-generation VLBI systems can be accommodated either by high-speed internet connections \citep{ngvla,ngVLA_update}, or increased capacity of hard disk and solid-state disk, which can be shipped by commercial carriers.  Once gathered at a central computing facility, the many 10's of Petabytes anticipated for next-generation EHT array observations can be correlated by purpose-built clusters, allocated time on national super-computing centers, or through virtual machine creation using cloud architectures (e.g., \citet{Gill2019}).  Once correlated, data analysis options include a growing number of video reconstruction algorithms that can render the dynamics of supermassive black hole activity on horizon scales \citep{Johnson2017,Bouman2018,Arras2022}.  These developments, combined with a positive mention of the ngEHT project \citep{Doeleman_2019} in the Radio/Millimeter/sub-Milimeter panel of the most recent US Astronomy Decadal Review \citep[][]{Astro2020}, imply that implementing the ngEHT is both feasible and timely.

\section{ngEHT Concept} \label{sec:concept}

The first images of M87 and Sgr~A* revealed a clear ring morphology, but they achieved a dynamic range of only ${\sim}10$ \citep{Paper4,sgr_paper3}.  Image fidelity from the 2017 data sets was primarily limited by sparse interferometric baseline coverage.  The shortest baselines, between telescopes located at the same geographic location (ALMA-APEX in Chile and JCMT-SMA in Hawai'i), probe arc second scale structures.  There is a large gap between these ``intra-site” baselines and the first ``inter-site” baseline that links LMT-SMT, which creates a baseline with angular resolution corresponding to $\sim 150\,\mu$as.  Furthermore, the 2017 observations included inter-site baselines between only five geographic locations for M87, and six locations for Sgr~A*, fundamentally limiting the fidelity of image reconstruction on angular scales that resolve the black hole shadow.  

\begin{figure}
    \centering
    \includegraphics[angle=0,origin=c,width=1.0\columnwidth]{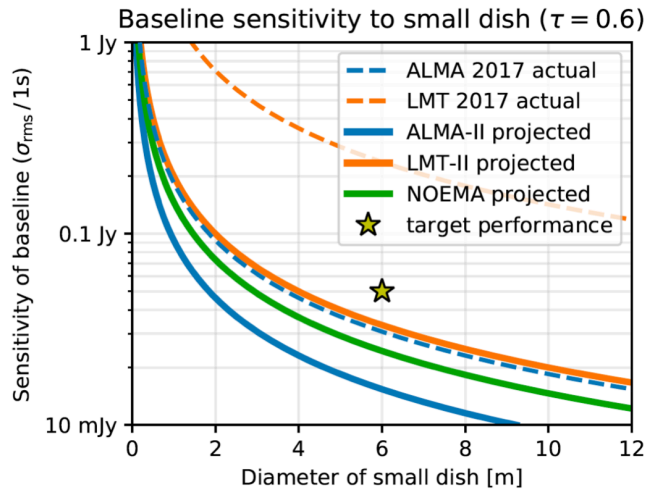}
    \caption{Interferometric baselines between key anchor stations and modest-diameter dishes have sufficient sensitivity to detect target flux densities on time-scales of several seconds.  
A star marks the correlated flux expected for SgrA* and M87
over long ngEHT baselines. Performance for 2017
is taken over 2 GHz of bandwidth and the observed
median sensitivity of ALMA and LMT during EHT
April 2017 observations. ALMA-II assumes phase referencing using the entire 8\,GHz (64\,Gbps) of EHT bandwidth, while LMT-II assumes 16\,GHz of bandwidth and aperture efficiency of $\eta_A=0.37$. NOEMA is projected for a 12-element array under nominal weather conditions,
and the small ngEHT remote site is evaluated at $\eta_A=0.5$ and line-of-sight opacity $\tau=0.6$. Atmospheric
phase tracked on rapid timescales at 86~GHz or 230~GHz can be transferred to 345~GHz, allowing for longer coherent integration times and robust measurement at the highest ngEHT observing frequencies.}
    \label{fig:anchor_plot}
\end{figure}

\begin{figure*}[t]
    \centering
    \includegraphics[width=1.00\textwidth]{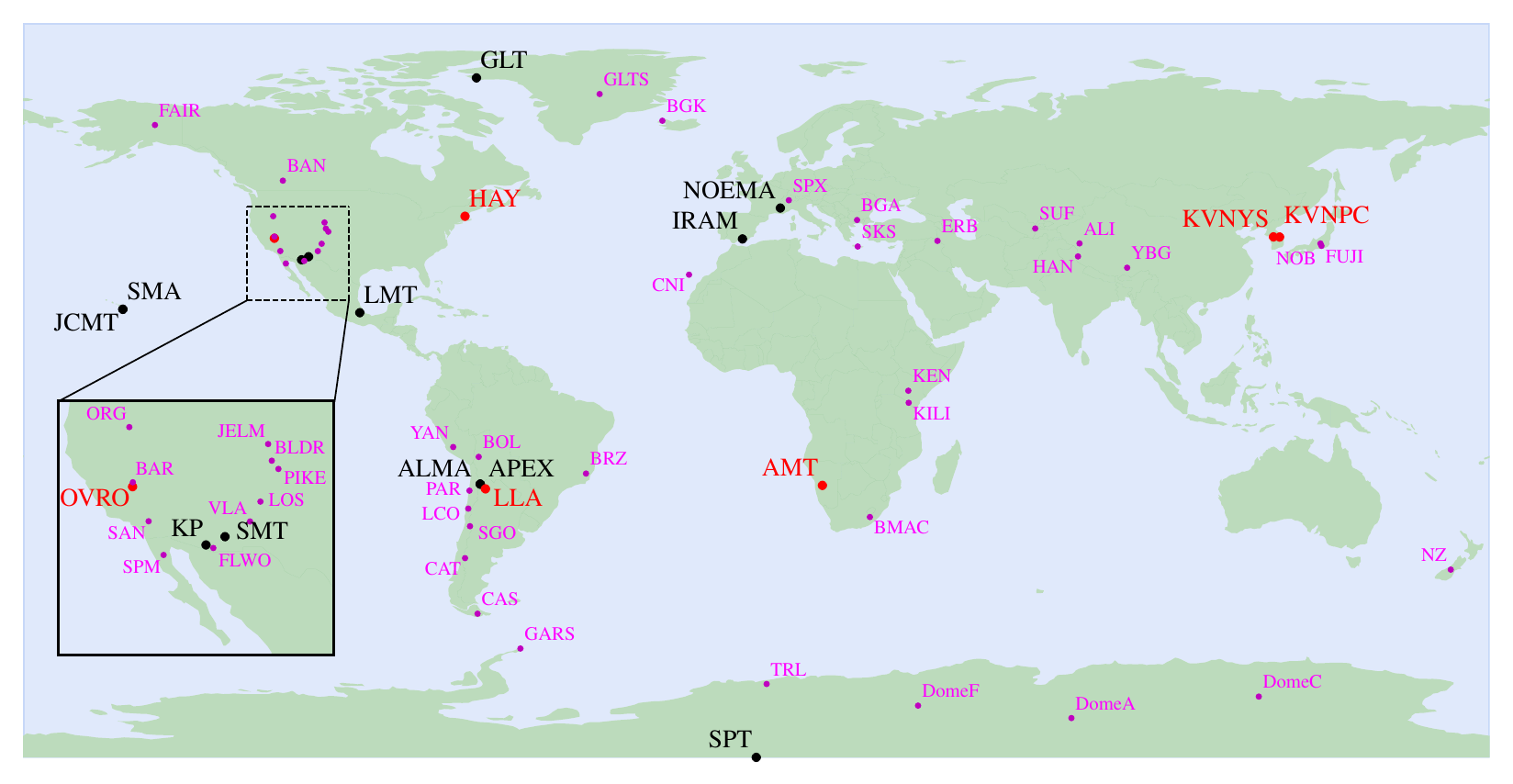}
    \caption{Current EHT sites (in black), other existing or near-future sites that may join global observations (in red), and potential new ngEHT sites (in magenta).}
    \label{fig:ngEHT_potential_sites}
\end{figure*}

The ngEHT concept focuses on overcoming these limits through several key developments.  Foremost among these is the deployment of relatively modest diameter radio dishes at optimized locations to increase baseline coverage.  \autoref{fig:anchor_plot} shows that even a 6m diameter dish in marginal weather conditions can detect long baseline correlated fluxes from Sgr~A* and M87 when paired with a large ``anchor” aperture.  This reflects the fact that the 2017 observations, though using fringe detection algorithms limited to 2GHz bandwidth, achieved signal-to-noise ratios that were typically in excess of $\sim 10$ and often reached $\sim 100$.  In other words, the current EHT is limited by baseline coverage and not sensitivity considerations.  Through this increased baseline coverage, the ngEHT will reach image dynamic ranges that exceed 1000:1 for full Earth rotation aperture synthesis observations of M87 and other AGN.  Time lapse movies that capture the dynamics of M87's accretion flow and jet launch by combining bi-weekly observations will achieve similar dynamic ranges (pre-cursor multi-epoch observations are possible with the existing EHT at lower imaging dynamic range).  For Sgr~A*, which has an Innermost Stable Circular Orbit (ISCO) period of $\sim \frac{1}{2}\,$hour, the ngEHT snapshot baseline coverage in 5 minute integrations will be sufficient for near real-time video reconstruction.  \autoref{fig:ngEHT_potential_sites} shows the current EHT array and the location of potential new ngEHT sites.

Significant improvements in sensitivity will still be realized through deployment of wider band receivers and backends, which can now typically digitize 8GHz per sideband.  For a given frequency band, the ngEHT targets dual-sideband and dual-polarization, for a potential Stokes I fringe detection that combines 32GHz aggregate bandwidth.  For any given baseline, this advance results in a net detection threshold that is four times lower than currently achievable.  

In addition to this increase in overall received bandwidth, the ngEHT frequency coverage will include the 86 and 345~GHz bands.  Routine multi-band operation has several important consequences for ngEHT capability.  Each station pair probes distinct spatial frequencies when observing in different bands, and multi-frequency imaging algorithms can make use of the aggregate interferometric coverage to improve image fidelity \citep[e.g.,][]{Chael_2022}.  Observing in the 345GHz band also improves angular resolution of the global array by up to 50\%.  The EHT already offers 345GHz observing capability on a subset of antennas \citep[][see also \autoref{tab:Array}]{Crew_345_2023}, but not yet simultaneously with 230GHz.  Additional frequency bands also enable analyses and modeling that differentiate between gravitationally lensed achromatic features (e.g., the photon ring) and structures whose appearance have a spectral dependence (e.g., accretion flows and relativistic jets).  And through use of the frequency phase transfer technique \citep[FPT;][]{Rioja_galaxies_2023}, VLBI phase solutions determined at lower frequencies can be transferred to higher frequency observations, effectively removing atmospheric phase effects to extend coherent integration times for higher sensitivity.  The full case for adding 86~GHz capability that leverages FPT through simultaneous multi-band systems is described in \citet{Issaoun_2023_86GHz}.   

Combined, these enhancements lead to profound increases in array capability.  The implementation roadmap for the ngEHT will proceed in two phases with a goal of ultimately adding $\sim10$ new dishes to the EHT.  In \phaseone, a total of six new sites will be developed: four radio dishes will be deployed to new geographic locations (\autoref{sec:SiteSelectionPhase1}); and two existing facilities (the 37m telescope at MIT Haystack Observatory and a 10m telescope at the Owens Valley Radio Observatory) will be modified to participate in future observations \citep[see, e.g.,][]{Kauffmann_2023}.  A \phasetwo will add four more telescopes, either by deploying additional new purpose-built telescopes, or by instrumenting planned single dish facilities due to come on-line by $\sim2030$ (\autoref{sec:SiteSelectionPhase2}).  These Phases, when complete, will double the number of dishes in the array recently fielded in the 2022 and 2023 annual EHT observing campaigns (see \autoref{sec:SiteSelection}).

\section{Next Generation Science Goals} \label{sec:Science}

The ngEHT design has been guided by a series of Key Science Goals (KSGs), developed through a community-driven process of exploration, evaluation, and prioritization. These goals and their associated instrument requirements are presented via a Science Traceability Matrix (STM) in a companion paper~\citep{ngEHT_KSG}; and a series of papers in a special issue of {\it Galaxies}\footnote{\url{https://www.mdpi.com/journal/galaxies/special_issues/ngEHT_blackholes}} presents science topics in greater detail.  Here, we briefly describe several of the ngEHT KSGs, which define the target baseline array architecture, and are summarized in Table~\ref{tab:science_goals}.

\subsection{Existence and properties of black hole horizons} \label{sec:existence}

By characterizing the central brightness depression region in black hole images, the ngEHT can directly address the question of the existence of a black hole's horizon. 
For Magnetically Arrested Disk (MAD) accretion modes, which are favored for M87 \citep{Paper5}, emission in the innermost part of the flow originates primarily in the equatorial plane, and the central depression (the ``inner shadow") is defined by light paths that cross the event horizon without visiting the emitting region of the accretion system \citep{Dokuchaev_2019, Chael_innershadow_2021}. Measuring the shape of this ``inner shadow" to be smaller than the photon orbit would correspond to observing the lensed event horizon, allowing estimates of the black hole's mass and spin \citep{Chael_innershadow_2021}. For both M87 and Sgr~A*, this measurement requires an imaging dynamic range of ${\sim}100$:1. For Sgr~A*, intrinsic variability presents an additional challenge that will require future algorithm development. Furthermore, enhancing the dynamic range of the images with the ngEHT will allow to obtain improved constraints on the brightness ratio between the black hole shadow interior and the observed emission ring. These constraints can be translated to an argument supporting the existence of the event horizon, by putting most stringent limits on the albedo of the surface of an exotic compact object alternative to a black hole \citep{sgr_paper6}, ultimately limited only by the emission and absorption in the foreground by the gas located out of the equatorial plane \citep[e.g.,][]{Vincent2022}.

\subsection{Measurements of the Spin of a SMBH}
\label{subsec:no-hair}

General relativity predicts that astrophysical black holes are described solely by two properties: their mass and angular momentum (or ``spin''). The ngEHT has the opportunity to produce the direct secure measurements of a black hole spin through distinctive image features that reflect the imprint of the strongly curved spacetime near the horizon. In particular, images of GRMHD simulations show several robust indicators of spin \citep[for a review, see][]{Ricarte_2022a}. The most promising of these is the spiraling polarization pattern around the emission ring \citep{Palumbo_2020}. By producing time-averaged polarimetric images of M87 and Sgr~A* at both 230 and 345\,GHz, the ngEHT will be able to securely measure this pattern and decouple the effects of the spacetime from those of the surrounding plasma (Faraday rotation and conversion), which are steeply chromatic.

\subsection{Evolution of Supermassive Black Holes.}

Though the EHT has to date observed only two SMBHs (\m87 and \sgra) with horizon-scale angular resolution, numerical simulations of black hole accretion flows \citep[e.g.,][]{Paper5,sgr_paper5} predict that the ``shadow'' structure seen towards these sources should be a generic image feature in sufficiently optically thin systems.  Measurements of the size of the SMBH shadow can be used to constrain the black hole mass  \citep[e.g.,][]{Paper6,sgr_paper4}, and measurements of the near-horizon linear polarization structure may also be able to provide indirect constraints on the black hole spin \citep{Palumbo_2020,Ricarte_2022a,Ricarte_2022b}.  Access to a population of shadow-resolved SMBHs would thus provide an opportunity to make uniquely self-consistent measurements of these spacetime properties, permitting corresponding studies of SMBH formation, growth, and co-evolution with their host galaxies.

The ngEHT is expected to be able to be able to detect up to several dozen SMBHs with sufficient angular resolution and sensitivity to access their masses and spins through measurements of their horizon-scale structure \citep{Pesce_holeographics_2021, Pesce_2022}.  A database of the most promising individual targets is being compiled within the ETHER sample \citep{Ramakrishnan_2023}.

\subsection{Mechanisms of Black Hole Accretion}

Despite decades of study, the mechanisms that drive accretion onto SMBHs are still poorly understood \citep[for a review, see][]{Yuan_Narayan_2014}. The ngEHT will make the first resolved movies of a black hole accretion flow, allowing a direct study of the dynamics of the turbulent plasma and the role of magnetic fields in providing an effective viscosity that drives infall \citep{Balbus_1991,Balbus_1998}.  

\subsection{Heating and Acceleration of Relativistic Electrons}
In low density, low accretion rate systems such as in M87$^\star$ and Sgr~A$^\star$ the Coulomb collision time for both electrons and protons is much larger than the dynamical (accretion) time scale. As a consequence, protons and electrons cannot redistribute their energy and a two-temperature plasma is occurs. Assuming that the emission is mainly generated by electrons, their temperature is determined by the interplay between cooling and heating processes \citep{Mahadevan_1997}. Given, that in low accretion rate systems the cooling processes can be neglected (cooling time scale larger than the dynamical time scale), the impact of possible electron heating processes on the observed emission from M87$^\star$ and Sgr~A$^\star$ can be probed. 

Two of the main processes for the heating of electrons are turbulent heating \citep[see, e.g.,][]{Kawazura2019,Howes2010}) and magnetic reconnection heating \citep[see, e.g.,][]{Rowan2017}. The results of two-temperature general relativistic magneto-hydrodynamic (GRMHD) simulations showed that magnetic reconnection heating leads to a disk dominated emission structure while turbulent heating tends to a disk-jet structure \citep{Ryan2018,Chael2018,Chael2019,Mizuno2021}.  The ngEHT with its improved u-v coverage and increased sensitivity will allow us to image and track at the same time the disk and faint jet structures on scales of 100\,M. Together with the multi-frequency capabilities of the ngEHT movies of the total intensity and the spectral evolution can be produced. These movies, in close combination with detailed numerical simulations, can allow us to locate the heating sites and  distinguish between the different electron heating process in M87$^\star$ and Sgr~A$^\star$.

\subsection{Energy Extraction from Black Holes}

Energy from a spinning black hole can be extracted via the Blandford-Znajek (BZ) process \citep{Blandford1977}, an electromagnetic analog of the classic Penrose process \citep{Penrose_1969}. With the ngEHT we will probe this energy extraction mechanism via the generated jet power or more precisely via the so-called BZ jet power. The BZ-jet power is proportional to the square of the black hole spin and to square of the magnetic flux crossing the horizon \citep{Tchekhovskoy2011}. In addition, the jet power can be measured from the observed spectral energy distribution or from the x-ray luminosity \citep[see][for a detailed disccusion on jet power estimates]{Prieto2016}. 

To compute a theoretical estimate for the BZ-jet power precise measurements of the black hole spin and the magnetic flux are necessary. As mentioned in \autoref{subsec:no-hair} of this paper and in \citet{Broderick_photonring_2022} combined ngEHT observations will provide the black hole spin and black hole mass with sufficient precision. The second quantity in the BZ-jet power, namely the magnetic flux across the horizon can be obtained either via polarimetric ngEHT observations \citep{EHT2021pol} or via the frequency dependent position of the core i.e., the core-shift using multi-frequency observations \citep{Zamaninasab2014}. In both cases the superior detection and imaging capabilities of the ngEHT will allow us to provide answers to this long-standing question of energy extraction from black holes. To perform this measurement for M87, \phasetwo of the ngEHT is required.

\subsection{Jet Formation}
Based on numerical GRMHD simulations we know that rotating black holes can launch jets via the BZ process \citep[see, e.g.,][]{Gammie2003,Tchekhovskoy2011}. However, once launched the jets need to be accelerated and confined, whereas the associated physical processes behind this acceleration and collimation as well as the jet composition are still on debate \citep[see][for a review]{Blandford2019}. The jet composition electron-positron or electron-proton plasma can be probed via circular polarisation and a detailed review can be found in this special issue by \citet{Emami2022}.

Details on the fluid structure and the formation process of the jet in M87 can be derived from the velocity field and the jet-to-counter-jet ratio \citep{Mertens2016,Walker2018}. The structure of the velocity field will allow us to probe the stratification of the jet into a fast inner spine and slow outer sheath \citep[see, e.g.,][]{Komissarov1999}. The ngEHT will enable such studies in objects other than M87 through resolving the transversal jet structure, e.g., in Centaurus A \citep{Janssen_2021}. In addition to the poloidal velocity field (spine-sheath structure), the toroidal velocity field plays a crucial role in determining the formation process of the jet: Is the jet anchored in the accretion disk \citep{Blandford1982} or is the jet launched from the ergosphere of a rotating black hole \citep{Blandford1977}. Extracting the velocity field of the jet requires multi-frequency observations and a high cadence of observations. To avoid the ``contamination'' of the velocity field by secondary effects, i.e. by Kelvin-Helmholtz instabilities or re-collimation shocks (triggered by changes in the ambient medium) scales up to 100\,M are sufficient. In order to determine the velocity field in M87 the ngEHT in \phaseone is required. 

\subsection{Constraining Properties of the Black Hole Photon Ring} \label{sec:photon_ring_science}
One of the clearest predictions motivated by the first black hole images is that the observed ring of emission should exhibit a fine sub-structure: nested concentric rings, each formed by light rays that make successively more orbits around the photon shell region, located very close to the black hole's event horizon \citep{Johnson_2020}. Each sub-ring is a lensed image of the surrounding accretion and jet emission with inner sub-rings becoming exponentially fainter and narrower. The structure of the primary ($n=0$) ring, observed by the EHT, depends on a combination of the local spacetime and the detailed emission structure on Schwarzschild radius scales, while subsequent sub-rings ($n\geq1$) asymptotically approach the true photon orbit, which is dependent exclusively on the spacetime metric \citep{Bardeen_1973}. Detection of the $n=1$ ring, formed by photons that make a half-orbit around the black hole, would be important confirmation of this untested prediction of General Relativity and lead to new tests of GR in highly curved space-time \citep{Broderick_photonring_2022,Wielgus2021}.  Robust extraction of this feature with the ngEHT will require the longest Earth baselines at 345~GHz and geometric model fitting that uses multiple frequencies \citep{Tiede_PhotonRing}.  This science goal would be a target of the fully realized (\phasetwo) ngEHT.

\begin{table*}
\small
     \begin{tabular}{ccccc}
     \hline
 {\bf Key Science Goal} &  {\bf Source} & {\bf ngEHT Phase} & {\bf Reference}\\
       \hline
       \hline
 Establish the existence & \m87   & \phaseone  & {\scriptsize \citet{Chael_innershadow_2021};} \\
 of black hole horizons  & \sgra  & \phasetwo & {\scriptsize \citet{Dokuchaev_2019}}  \\ \hline
 Measure a SMBH's spin   & \m87   & \phasetwo & {\scriptsize \citet{Palumbo_2020}} \\
                         & \sgra  & \phasetwo & {\scriptsize \citet{Ricarte_2022a}} \\ \hline 
 Understanding Black Hole-Galaxy   & AGN Survey & \phaseone & {\scriptsize \citet{Pesce_holeographics_2021,Pesce_2022};} \\
 Formation, Growth and Coevolution &  & & {\scriptsize \citet{Ramakrishnan_2023}} \\ \hline
 Reveal how black holes  & \m87  & \phaseone & {\scriptsize \citet{Balbus_1998};} \\
 accrete material        & \sgra & \phasetwo & {\scriptsize \citet{Yuan_Narayan_2014}} \\ \hline
 Observe localized electron & \m87 & \phaseone & {\scriptsize  \citet{Rowan2017};}\\
 heating and acceleration  & \sgra & \phasetwo & {\scriptsize  \citet{Ball_2018}}\\ \hline
 Determine if BH jets are & \m87  & \phasetwo & {\scriptsize \citet{Blandford1977};} \\
 powered by spin energy  & \sgra & \phasetwo & {\scriptsize \citet{Tchekhovskoy2011}} \\ \hline
 Determine jet formation & \m87  & \phaseone & {\scriptsize \citet{Blandford2019}} \\
 \& launching mechanisms & \sgra & \phasetwo & \\ \hline 
 Constraining Properties & \m87  & \phasetwo & {\scriptsize \citet{Johnson_2020};} \\
 of the BH Photon Ring  & \sgra & \phasetwo & {\scriptsize \citet{Tiede_PhotonRing}} \\ \hline \vspace{-0.5cm}
   \end{tabular}
\caption{Select ngEHT Key Science Goals. For the full Science Traceability Matrix and additional details, see \citet{ngEHT_KSG}.
}
 \label{tab:science_goals}
 \end{table*}

\section{Optimizing the ngEHT Reference Array} \label{sec:OptimizingArray}

The scientific performance of an array generically benefits from addition of new stations, regardless of where those stations are located.  However, when constrained by a fixed budget or a fixed number of new dishes to be added, determining the optimal placement of the new dishes is a challenge that requires finding a balance between many -- often conflicting -- objectives.  For instance, science goals that require high angular resolution favor array configurations with many long baselines, while goals that involve high-fidelity imaging on large fields of view instead favor configurations containing dense short-baseline coverage.  Similarly, while atmospheric opacity considerations favor the highest and driest locations, such sites are often remote and lack critical infrastructure, significantly driving up construction and operating costs.  Any array configuration that one ultimately arrives at necessarily hinges on a non-unique choice about what exactly constitutes ``optimality,'' and the result can depend sensitively on how one weights the many relevant considerations when doing so.

In this section, we detail some of the considerations that are entering into the design process for the ngEHT array configuration.  \autoref{sec:CandidateSites} describes how we have selected an initial pool of candidate sites to consider, \autoref{sec:SyntheticData} describes our procedure for simulating realistic ngEHT observations, and \autoref{sec:Metrics} details several metrics that we use to evaluate array quality.  \autoref{sec:SiteSelection} describes our evaluation of the many different candidate arrays and discusses a strategy for translating array performance into site selection.  Various details relevant for the site selection procedure are provided in \autoref{app:SiteDetails}.

\subsection{Candidate sites} \label{sec:CandidateSites}

From most locations on the surface of the Earth, atmospheric opacity prevents observations at the primary ngEHT frequencies of 230\,GHz and 345\,GHz.  We thus take our starting pool of candidate sites from \cite{Raymond_2021}, who identified sites with favorable atmospheric transmission properties for 230\,GHz and 345\,GHz observations during the March/April typical EHT observing season; the candidate sites are shown in \autoref{fig:ngEHT_potential_sites} and listed in \autoref{tab:CandidateSites}.

\begin{figure}
    \centering
    \includegraphics[width=1.00\columnwidth]{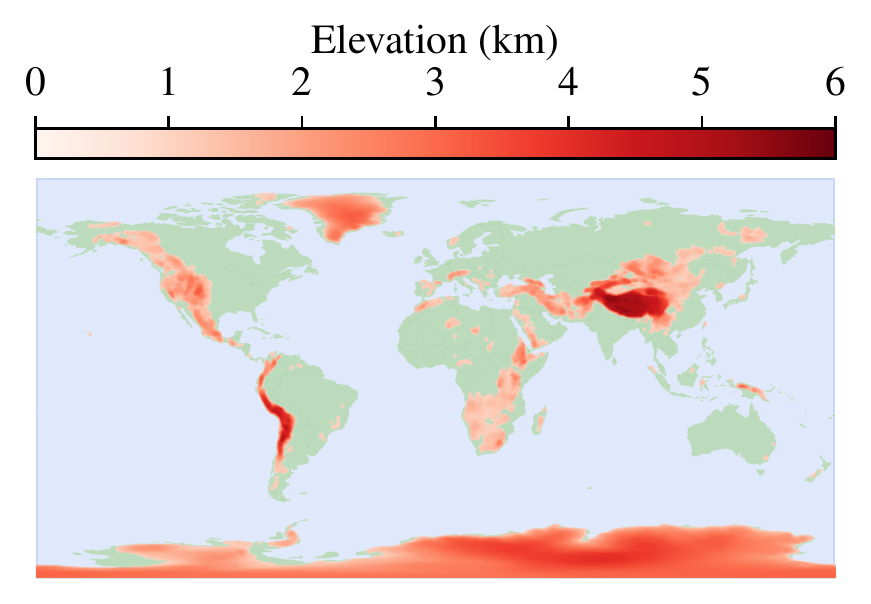}
    \caption{Global elevation map.  Locations with elevations above 1000 meters are shaded red, with darker colors indicating higher elevations.}
    \label{fig:global_elevation_plot}
\end{figure}

\begin{figure}
    \centering
    \includegraphics[width=1.0\columnwidth]{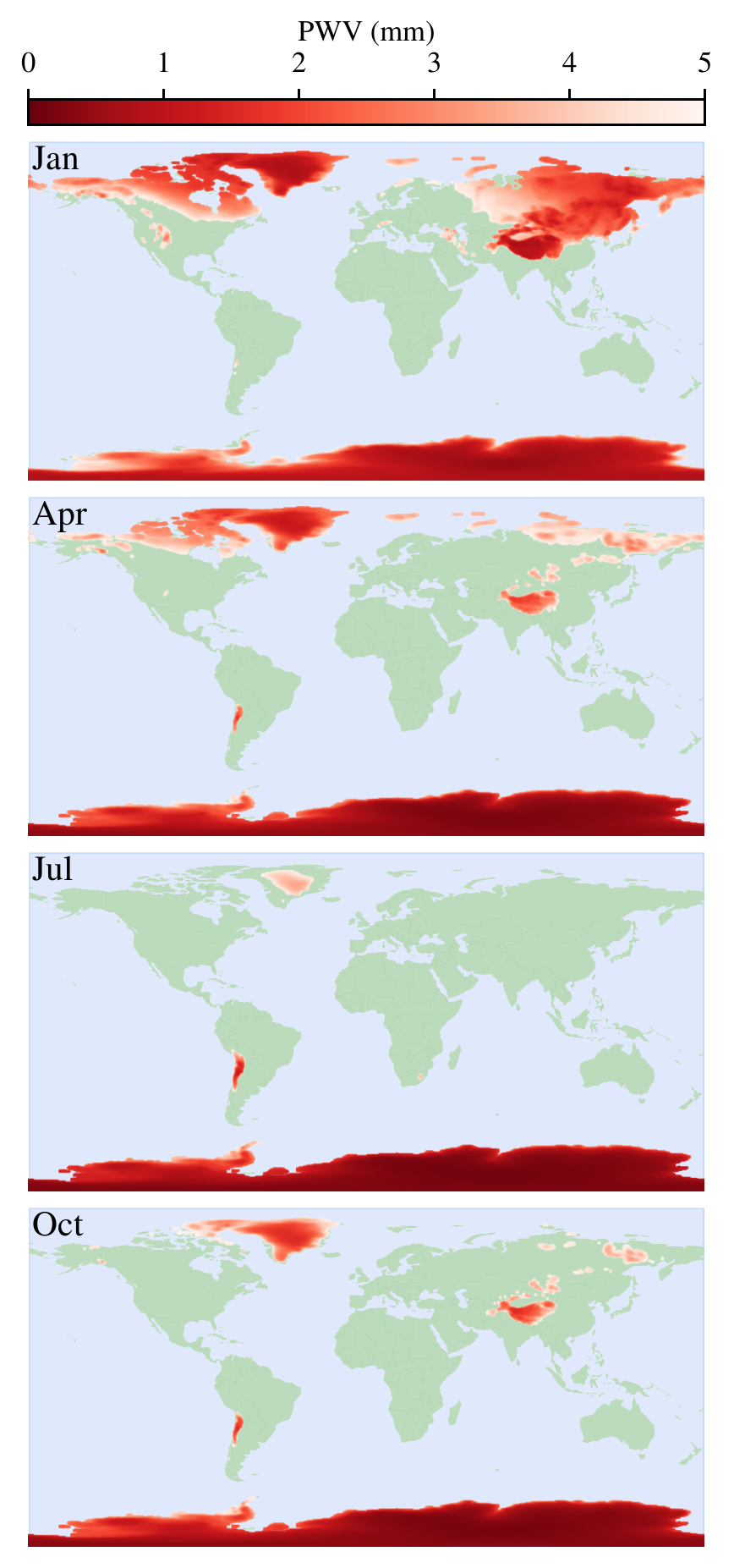}
    \caption{Locations around the globe with mean PWV less than 5\,mm, in January (top), April (second from top), July (second from bottom), and October (bottom).  A darker red coloring indicates a lower value of mean PWV, and only locations with elevations above 50 meters are colored.  We have determined the PWV via \autoref{eqn:PWV} using atmospheric data from MERRA-2 \citep{MERRA2}, and the average is taken over all available data between 2012--2022.}
    \label{fig:global_PWV_plot}
\end{figure}

Given the selection on atmospheric opacity performed in \cite{Raymond_2021}, the candidate sites are naturally situated in the highest and driest locations. \autoref{fig:global_elevation_plot} shows the highest-elevation locations around the globe, and \autoref{fig:global_PWV_plot} shows where the mean level of precipitable water vapor (PWV) is lowest throughout the year.  We have computed the PWV using atmospheric data from the MERRA-2 database \citep{MERRA2}.  The PWV at a particular location is determined by integrating the water vapor through the column of atmosphere above that location \citep[see, e.g.,][]{Salby_1996},
\begin{equation}
    \text{PWV} = \frac{1}{\rho g} \int_0^{P_{\text{surf}}} \frac{q(P)}{1-q(P)} dP . \label{eqn:PWV}
\end{equation}
\noindent Here, $q(P)$ is the specific humidity, $P$ is the atmospheric pressure, $P_{\text{surf}}$ is the atmospheric pressure at the surface, $\rho \approx 1$\,g\,cm$^{-3}$ is the mass density of water, and $g \approx 9.81$\,m\,s$^{-2}$ is the acceleration of gravity at the surface of the Earth.  MERRA-2 provides both $P$ and $q$ in 42 different atmospheric layers as a function of geographic location and time.

\subsection{Synthetic data generation} \label{sec:SyntheticData}

We evaluate candidate array performance using synthetic observations of the key science targets \m87 and \sgra.  For source models we use the results of general relativistic magnetohydrodynamic (GRMHD) simulations that have been post-processed using ray-tracing and radiative transfer codes to produce images at the 230\,GHz and 345\,GHz observing frequencies appropriate for the ngEHT.  Our \m87 source model comes from the simulations carried out in \cite{chael_2019}, and our \sgra source model comes from the simulation library produced in \cite{sgr_paper5}.

We generate synthetic datasets using the \texttt{ngehtsim}\footnote{\url{https://github.com/Smithsonian/ngehtsim}} library.  Given a candidate ngEHT array configuration and a source model, \texttt{ngehtsim} uses \texttt{eht-imaging} \citep{Chael_2016,Chael_2018} to sample the Fourier transform of the source at the \uv-coverage corresponding to the array.  Thermal noise $\sigma_{ij}$ on a baseline between stations $i$ and $j$ is determined by the radiometer equation,
\begin{equation}
    \sigma_{ij} = \frac{1}{\eta_q} \sqrt{\frac{\text{SEFD}_i \text{SEFD}_j}{2 \Delta \nu \Delta t}} , \label{eqn:ThermalNoise}
\end{equation}
\noindent where $\Delta \nu$ is the observing bandwidth, $\Delta t$ is the integration time, SEFD is the station system equivalent flux density, and $\eta_q = 0.88$ is an efficiency factor associated with 2-bit quantization during data collection \citep{TMS}.  We determine SEFDs for each station as a function of time using
\begin{equation}
    \text{SEFD} = \frac{2 k T_\text{sys}}{A_\text{eff}} e^{\tau} , \label{eqn:SEFD}
\end{equation}
\noindent where $k$ is the Boltzmann constant, $\tau$ is the (time-dependent) line-of-sight atmospheric opacity, $A_\text{eff}$ is the effective collecting area of the telescope,
\begin{equation}
    T_{\text{sys}} = T_{\text{rx}} + T_{\text{atm}} \left( 1 - e^{-\tau} \right) \label{eqn:Tsys}
\end{equation}
\noindent is the system temperature, $T_{\text{rx}}$ is the receiver temperature, and $T_{\text{atm}}$ is the temperature of the atmosphere.  We determine $T_{\text{atm}}$ using historical atmospheric data from the MERRA-2 database \citep{MERRA2}, and $\tau$ is obtained by passing the atmospheric state information from MERRA-2 through the \textit{am} radiative transfer code \citep{Paine_2019}.

For the synthetic datasets in this section, we use a bandwidth of $\Delta \nu = 16$\,GHz (for ngEHT) or $\Delta \nu = 2$\,GHz (for EHT) at each of two frequency bands, one centered at 230\,GHz and the other centered at 345\,GHz.  We use an integration time of $\Delta t = 10$\,minutes, which is assumed to be enabled by suitable phase calibration, with a 50\% duty cycle (i.e., 10 minutes on-source followed by 10 minutes off-source); the total duration of each observation is 24 hours.  We assume receiver temperatures $T_{\text{rx}}$ of 50\,K at 230\,GHz and 75\,K at 345\,GHz.

To emulate fringe-finding, we apply two signal-to-noise ratio (SNR) thresholding schemes to the generated visibilities.  The first scheme emulates the ``fringe groups'' strategy from \citet{ehthops}: if a visibility does not achieve an equivalent SNR of 5 on an integration time of 10 seconds (at 230\,GHz) or 5 seconds (at 345\,GHz), and if the stations comprising the baseline associated with that visibility do not participate in other baselines that achieve the requisite SNR, then that visibility is assumed to be a non-detection and is flagged from the dataset.  Note that for the stations with dual-frequency capabilities, both of the frequency bands are checked simultaneously; if either one of the two frequency bands has an SNR that satisfies the threshold condition, then we assume that both bands can be detected.  This second scheme emulates frequency phase transfer across the bands \citep[see, e.g.,][]{RiojaDodson2020}.  Both of these fringe-finding schemes are applied when simulating ngEHT data, while only the first (fringe groups) scheme is applied when simulating EHT data.

\subsection{Array performance metrics}\label{sec:Metrics}

The analysis methods utilized by the EHT for performing measurements of physical interest using VLBI data are in general highly computationally expensive to evaluate. Further, the added value of a particular set of new sites is non-linear in the number of sites; the number of new baselines is quadratic in the number of existing sites, and the value of an individual site is sensitive to its position with respect to existing dishes.

To evaluate the performance of candidate ngEHT array configurations without running computationally expensive analysis pipelines (such as imaging or model-fitting), we utilize metrics of array performance that are based on pre-analysis quantities. We primarily employ two metrics: one metric that quantifies the \uv-coverage and another metric that quantifies the aggregate baseline sensitivity.  We compute the array performance metrics using synthetic observations at frequencies of 230\,GHz and 345\,GHz, which drive the key science goals of the ngEHT.  While 86\,GHz is an important addition that enables improved detection prospects at the higher frequencies (see \autoref{sec:Receiver}), it serves primarily a calibration-related role and thus is not included in our \uv-coverage or baseline sensitivity metric computations.

\begin{figure}[t]
    \centering
    \includegraphics[width=1.0\columnwidth]{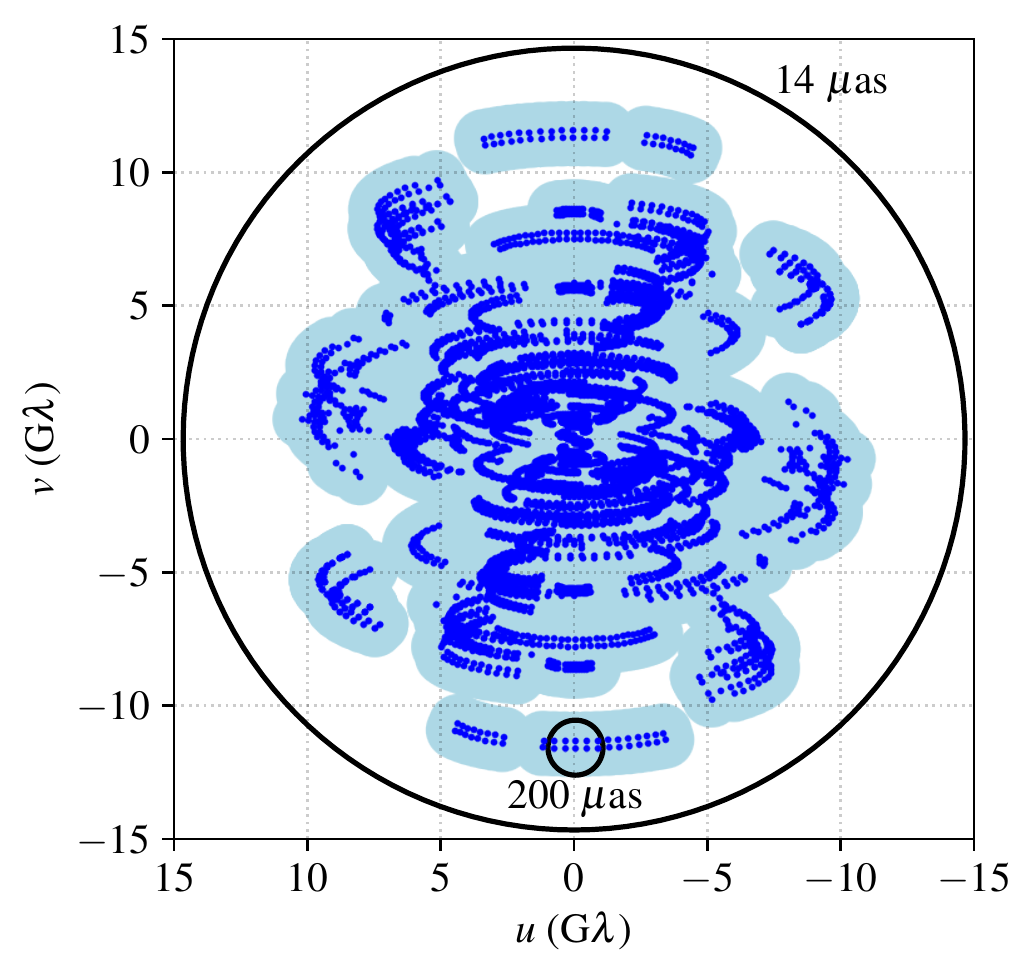}
    \caption{Illustration of the \uv-filling fraction metric described in \autoref{sec:Metrics} (see also \citet{Palumbo_2019}). Given some \uv-coverage -- shown here by the blue points for a mock observation of M87 using the full ngEHT \phaseone array in April -- the FF metric is a measure of how much area within a circular region of radius $1/\theta_{\text{res}}$ is sampled, after convolving the coverage with a disk of radius $1/\theta_{\text{FOV}}$.  In this case, $\theta_{\text{res}} = 14$\,\uas and $\theta_{\text{FOV}} = 200$\,\uas.  The convolved coverage is shaded in light blue, and takes up a fraction $\mu_{\text{ff}} = 0.5$ of the area of the outer circle.}
    \label{fig:FF_illustration}
\end{figure}

We use as our quantification of \uv-coverage quality the \uv-filling fraction metric (FF metric), defined in \citet{Palumbo_2019} as the fraction \ff of the area enclosed by a bounding circle in \uv of radius $1/\theta_{\text{res}}$ that is covered by the two-dimensional convolution of the coverage with a circular disk of radius $1/\theta_{\text{FOV}}$. Here, $\theta_{\text{res}}$ and $\theta_{\rm FOV}$ are array performance specifications based on imaging expectations, and they are not predicted directly by the coverage; \autoref{fig:FF_illustration} provides an illustration of how the FF metric is calculated.  \citet{Palumbo_2019} found that as the filling fraction increases, imaging performance in compact imaging examples improves steadily until it flattens to a constant factor of the diffraction-limited image fidelity near \ff $\gtrsim 0.5$. The FF metric naturally demands greater coverage for equivalent \ff as expectations of imaging field of view $\theta_{\rm FOV}$ increases; however, \ff does not capture the relative information density of the Fourier plane, and for many source morphologies, the importance of Fourier coverage decreases with radius from the \uv-coordinate origin.  In this paper, we assume $\theta_{\text{res}} = 14$\,\uas (i.e., the angular resolution of an Earth-diameter baseline observing at 345\,GHz) unless otherwise specified, but we use several different fields of view; when quoting FF metric values, we will thus specify the corresponding assumed of view using the notation $\mu_{\text{ff}}(\theta_{\text{FOV}})$.

For our quantification of array sensitivity, we use the point source sensitivity (PSS) metric,
\begin{equation}
\text{PSS} = \left( \sum_{i=1}^N \frac{1}{\sigma_i^2} \right)^{-1/2} ,
\end{equation}
\noindent where $\sigma_i$ is the value of the thermal noise on visibility $i$ (see \autoref{eqn:ThermalNoise}), and the sum is taken over all visibilities in the dataset.  The PSS metric, which has units of flux density, quantifies the sensitivity that the array could in principle achieve when measuring the flux density of a point source.  It naturally folds in not only the observing bandwidth and diameter of each telescope in the array, but also the amount of mutual visibility that each site has with every other as well as the atmospheric transmission at each site.

\subsection{Site selection} \label{sec:SiteSelection}

The stringent atmospheric opacity requirements for observing at millimeter wavelengths means that only a small number of locations around the globe are suitable candidates (see \autoref{sec:CandidateSites}).  Given that the list of candidate sites presents a finite number of discrete locations on the globe where telescopes could be placed, we could in principle evaluate all possible new array configurations.  The ability to confine the site search space to a finite number of options in this way is fairly unique to high-frequency VLBI, and it informs our optimization strategies below; the analogous site selection problem for connected-element arrays (e.g., VLA, ALMA, ngVLA) and low-frequency VLBI arrays (e.g., VLBA, SKA) presents a qualitatively different challenge.

In practice, though the number of possible new array configurations is finite, the space remains large and difficult to search comprehensively; the number of possible new array configurations that could be made using the 44 sites listed in \autoref{tab:CandidateSites} is approximately $1.8 \times 10^{13}$.  Additionally, we would like to ensure that the selected sites enable the ngEHT array to perform well across all of the following situations:
\begin{itemize}
	\item in observations of both \m87 and \sgra;
	\item during observations that take place throughout the year;
	\item when observing alongside any subset of the EHT.
\end{itemize}
The performance of each candidate array must also be evaluated using several different quality metrics (see \autoref{sec:Metrics}) that correspond to the various scientific goals.  All of the above considerations result in multiplicative factors that further increase the expense of a comprehensive analysis.

Given the difficulty of comprehensively searching all possible combinations of new stations, we instead partition our site selection efforts into two stages corresponding to the two anticipated phases of ngEHT development.  In the first stage -- corresponding to ngEHT \phaseone -- we consider the selection of three new sites from the pool of candidates.  The availability of three 6.1-meter BIMA dishes for refurbishment and relocation (see \autoref{sec:concept}) provides a pathway to realizing a \phaseone ngEHT array on a shorter ($\sim$few-year) timescale than it will take to field a larger array of newly-constructed dishes.  Optimizing for only three new sites at a time also reduces the number of site combinations to only ${44 \choose 3} = 13244$.  In the second stage of the site selection analysis -- corresponding to ngEHT \phasetwo -- we then consider the selection of five new sites from the remaining pool of candidates, corresponding to ${41 \choose 5} = 749398$ different site combinations.  Dividing the optimization strategy into two stages in this way, and selecting a specific target number of new sites in each stage, substantially reduces the computational cost of optimizing the array configuration.

The sites and frequency configurations corresponding to the selected \phaseone and \phasetwo ngEHT arrays are listed in \autoref{tab:Array}.

\begin{table*}
\centering
\begin{tabular}{l|c|c|c|c}
    \hline
     Site & \textbf{EHT} & \textbf{ngEHT \phaseone} & \textbf{ngEHT \phasetwo} & \textbf{ngEHT \phasetwo (alt.)} \\
    \hline
    ALMA & 86\hphantom{+}230\hphantom{+}345 & 86\hphantom{+}230\hphantom{+}345 & 86\hphantom{+}230\hphantom{+}345 & 86\hphantom{+}230\hphantom{+}345 \\
    AMT & - & - & - & 86+230+345 \\
    APEX & \hphantom{86+}230\hphantom{+}345 & 86\hphantom{+}230\hphantom{+}345 & 86\hphantom{+}230\hphantom{+}345 & 86\hphantom{+}230\hphantom{+}345 \\
    BOL & - & - & 86+230+345 & - \\
    CNI & - & 86+230+345 & 86+230+345 & 86+230+345 \\
    GLT & 86\hphantom{+}230\hphantom{+}345 & 86+230+345 & 86+230+345 & 86+230+345 \\
    HAY & - & 86+230\hphantom{+345} & 86+230\hphantom{+345} & 86+230\hphantom{+345} \\
    IRAM & 86+230\hphantom{+}345 & 86+230\hphantom{+}345 & 86+230\hphantom{+}345 & 86+230\hphantom{+}345 \\
    JCMT & 86\hphantom{+}230\hphantom{+}345 & 86+230+345 & 86+230+345 & 86+230+345 \\
    JELM & - & - & 86+230+345 & 86+230+345 \\
    KILI & - & - & 86+230+345 & - \\
    KP & 86\hphantom{+}230\hphantom{+345} & 86+230\hphantom{+345} & 86+230\hphantom{+345} & 86+230\hphantom{+345} \\
    KVNPC & - & - & - & 86+230+345 \\
    KVNYS & - & - & - & 86+230+345 \\
    LCO & - & 86+230+345 & 86+230+345 & 86+230+345 \\
    LLA & - & - & - & 86+230+345 \\
    LMT & 86\hphantom{+}230\hphantom{+345} & 86+230+345 & 86+230+345 & 86+230+345 \\
    NOEMA & 86\hphantom{+}230\hphantom{+}345 & 86+230\hphantom{+}345 & 86+230\hphantom{+}345 & 86+230\hphantom{+}345 \\
    OVRO & - & 86+230\hphantom{+345} & 86+230\hphantom{+345} & 86+230\hphantom{+345} \\
    SGO & - & - & 86+230+345 & - \\
    SMA & \hphantom{86+}230\hphantom{+}345 & \hphantom{86+}230\hphantom{+}345 & \hphantom{86+}230\hphantom{+}345 & \hphantom{86+}230\hphantom{+}345 \\
    SMT & \hphantom{86+}230\hphantom{+}345 & 86+230+345 & 86+230+345 & 86+230+345 \\
    SPM & - & 86+230+345 & 86+230+345 & 86+230+345 \\
    SPT & \hphantom{86+}230\hphantom{+345} & 86+230+345 & 86+230+345 & 86+230+345 \\
    SPX & - & - & 86+230+345 & - \\
    \hline
 \end{tabular}
 \caption{Site participation and frequency capabilities for the EHT and both phases of the ngEHT array.  For the first column, EHT sites with existing 86\,GHz capability are noted, but the EHT does not currently support 86\,GHz operation; and some of these 86\,GHz receivers cannot be used simultaneously with higher frequency receivers.  In each of the three rightmost columns, sites that do not participate in the specified array are indicated with a ``-'' sign.  Multi-frequency capabilities are indicated with a ``+'' sign; e.g., ``230+345'' indicates that the station can observe at both 230\,GHz and 345\,GHz simultaneously, whereas ``230 345'' indicates that it can only observe at each frequency separately.  For completeness, we list in the rightmost column an alternative incarnation of the ngEHT \phasetwo array, in which we forgo the need to field new telescopes by relying instead on external facilities that are anticipated to come online in the next few years (see \autoref{sec:AlternateStaging}).  For this alternate case, the JELM site would be added in Phase 1.} \label{tab:Array}
\end{table*}

\begin{figure*}[t]
    \centering
    \includegraphics[width=1.00\textwidth]{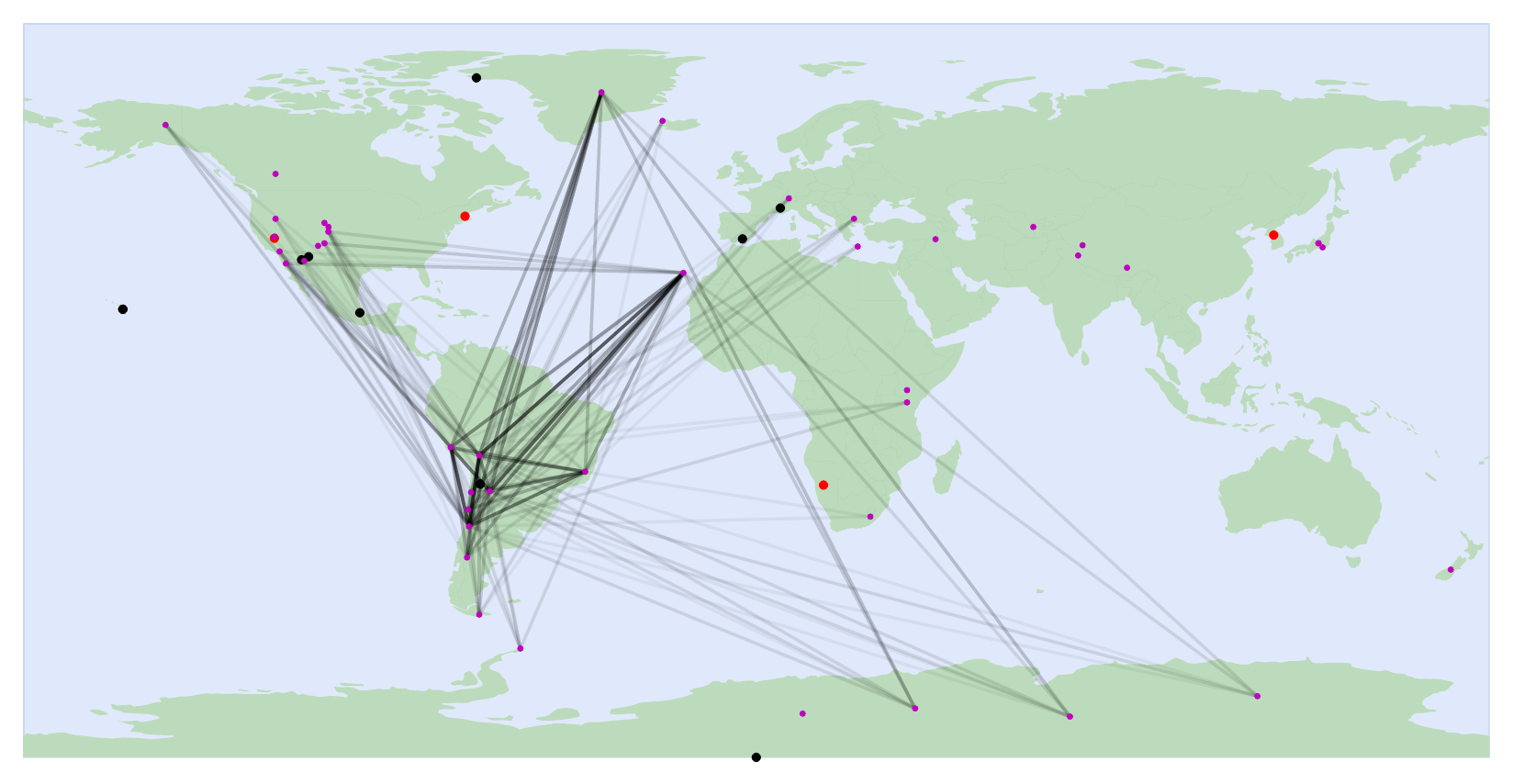}
    \caption{A version of the site map from \autoref{fig:ngEHT_potential_sites} with the top 1\% of all three-station candidate site combinations from the \phaseone exploration (see \autoref{sec:SiteSelection}) plotted as black triangles.}
    \label{fig:site_selection_map}
\end{figure*}

\subsubsection{\phaseone} \label{sec:SiteSelectionPhase1}

To determine the optimal locations for the three new \phaseone dishes, we carry out a survey of all possible three-station combinations of the 44 sites listed in \autoref{tab:CandidateSites}.  For each candidate set of three sites, we explore the performance of the resulting array (1) for observations of both \m87 and \sgra, (2) under weather conditions appropriate for January, April, July, and October, and (3) when observing alongside four different variants of the existing EHT array (specified in the top section of \autoref{tab:BaseArrays}).  These pre-existing array variants include various subsets of the EHT array, as well as the HAY and OVRO dishes that are expected to be outfitted with ngEHT equipment (see \autoref{sec:concept}).  We evaluate each candidate array using the metrics described in \autoref{sec:Metrics} for 100 instantiations of the weather conditions at each site, from which we then take median values to establish typical performance.

After evaluating all candidate arrays, we determine a ``performance score'' for each array according to its average ranking across the full suite of observing parameters.  E.g., if a particular array is ranked first for one set of observing parameters, ranked third for a second set of observing parameters, and ranked fifth for a third set of observing parameters, then its performance score would be $(1+3+5)/3 = 3$.  Arrays with smaller values of the performance score are those that have performed well across a range of observing parameters.  \autoref{fig:site_selection_map} shows the top 1\% of all three-station candidate site combinations after ranking them by their performance scores, with each set of sites plotted as a connected three-baseline triangle.  We identify six heavily populated clusters of high-performing site combinations:
\begin{itemize}
    \item An ``eastern cluster'' containing two sites in South America and either CNI or, less frequently, one of the other mainland European sites (BGA, SKS, SPX).
    \item A ``western cluster'' containing two sites in South America and a site in North America, most typically either SPM, PIKE, or FAIR.
    \item A ``northern cluster'' containing two sites in South America and GLTS, or less commonly with BGK.
    \item A ``southern cluster'' containing two sites in South America and one of the Antarctic Dome sites.
    \item An ``equatorial cluster'' containing one site in South America, one site in North America, and CNI.
    \item A ``polar cluster'' containing one site in South America, GLTS, and one of the Antarctic Dome sites.
\end{itemize}
\noindent We see that the most favored sites tend to be those that are able to leverage simultaneous observability with existing sites.  The overrepresentation of existing sites in the Western hemisphere means that sites in the Eastern hemisphere -- particularly those in Asia and New Zealand -- are correspondingly penalized.

To select from among the top-performing three-site combinations, we impose additional, more qualitative considerations.  We disfavor the northern, southern, and polar clusters because they contain sites that are unable to observe either \m87 (in the case of the Antarctic sites) or \sgra (in the case of GLTS and BGK).  The eastern and western clusters suffer from a similar asymmetry, in that they include sites that have little mutual visibility with existing American and European stations, respectively.  The equatorial cluster provides the most balance in terms of site geography, and it contains the three most favored regions for a new site: South America, North America, and CNI.  Several of the South American sites are comparably well-represented among the top site combination candidates, as are a couple of the North American sites.  After additionally accounting for initial site cost estimates and favoring lower-cost sites, we settle on the three-site combination of CNI, LCO, and SPM as our fiducial ngEHT \phaseone additions.

\subsubsection{\phasetwo} \label{sec:SiteSelectionPhase2}

In the second stage of our site selection analysis -- corresponding to ngEHT \phasetwo -- we consider the addition of five new sites to the previous three determined from the \phaseone selection.  We explore the same observing targets and weather conditions as for the \phaseone exploration, but we use updated pre-existing arrays that include the \phaseone sites (see the bottom section of \autoref{tab:BaseArrays}).  We again evaluate each candidate array using the metrics described in \autoref{sec:Metrics} for 100 instantiations of the weather conditions at each site, and we use median metric values to establish typical performance.

The selection process for \phasetwo is ongoing, but preliminary results indicate that the combination of BOL, JELM, KILI, SGO, and SPX would provide a strong improvement to the array coverage.  We thus take these sites to be our fiducial ngEHT \phasetwo additions for the purposes of this paper.

\subsubsection{Alternate Staging of New Sites} \label{sec:AlternateStaging}
Several new radio telescopes that could be used for ngEHT observations are planned to become operational in the coming years.  Thus, an alternative staging approach would be to augment \phaseone by adding JELM to the three new sites described in \autoref{sec:SiteSelectionPhase1}, and \phasetwo could then consist solely of the following planned telescopes: the LLAMA telescope in Argentina, the AMT in Namibia, the KVNYS telescope near Seoul, Korea, and the KVNPC telescope (currently under construction) near Pyeongchang, Korea.  Together, the four \phaseone sites (CNI, JELM, LCO, SPM) combined with OVRO, HAY, and the four planned telescopes (LLAMA, AMT, KVNYS, KVNPC) would constitute a near-doubling of the existing EHT array and would achieve comparable $(u,v)$-coverage to that provided by the array described in \autoref{sec:SiteSelectionPhase2}. This alternate pathway to a \phasetwo ngEHT would also provide capabilities sufficient to achieve all ngEHT Key Science Goals.

\subsubsection{Baseline Coverage}

\begin{figure*}[htbp!]
    \centering
    \includegraphics[width=1.00\textwidth]{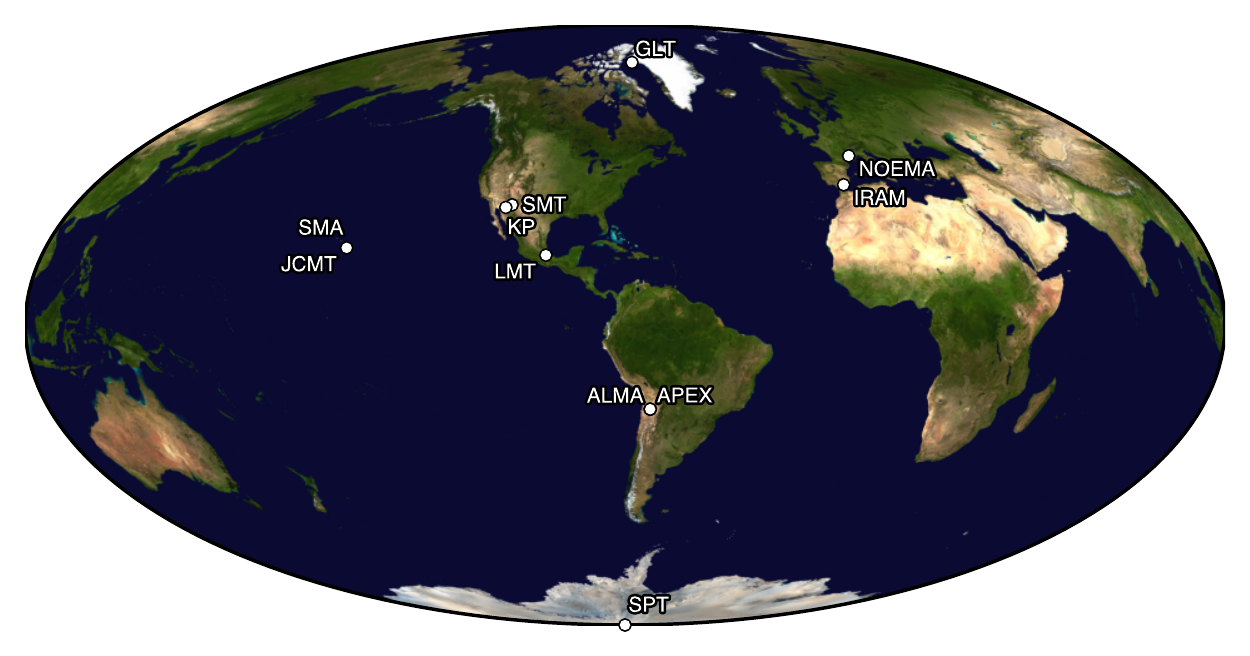}
    \includegraphics[width=0.49\textwidth]{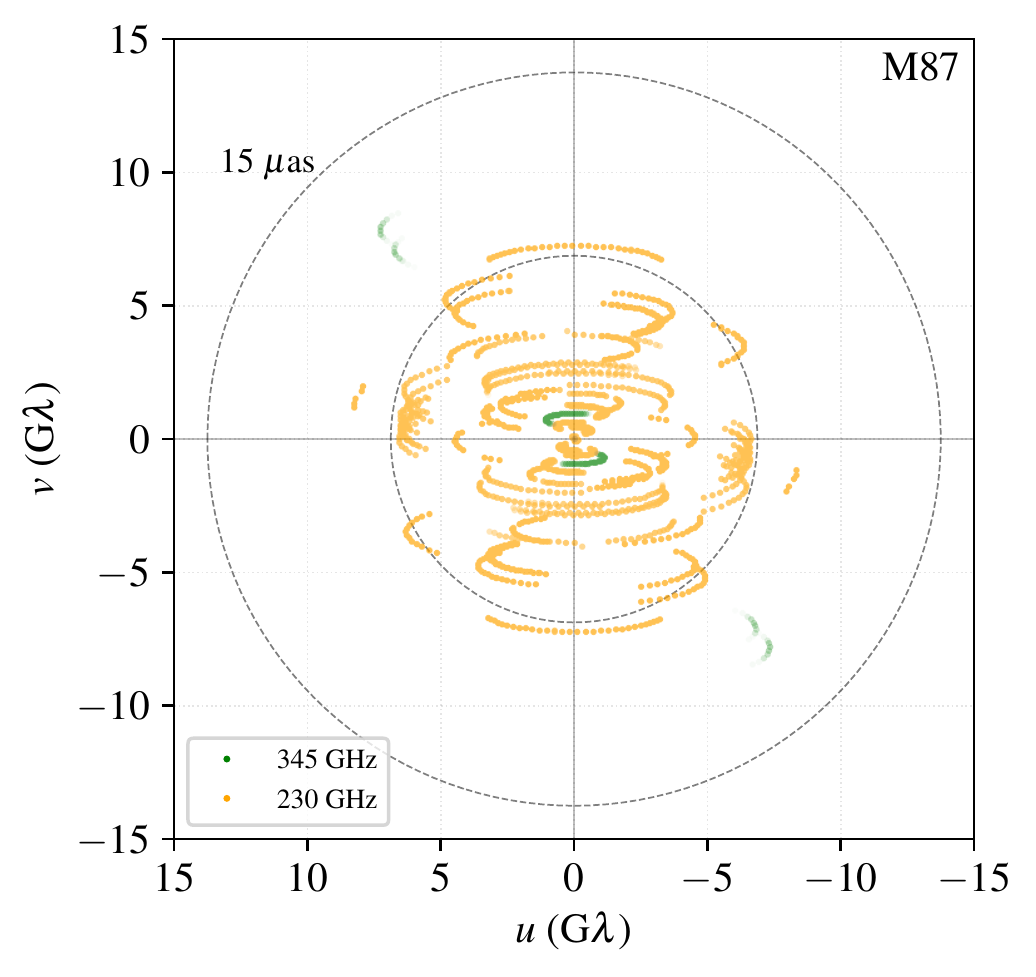}
    \includegraphics[width=0.49\textwidth]{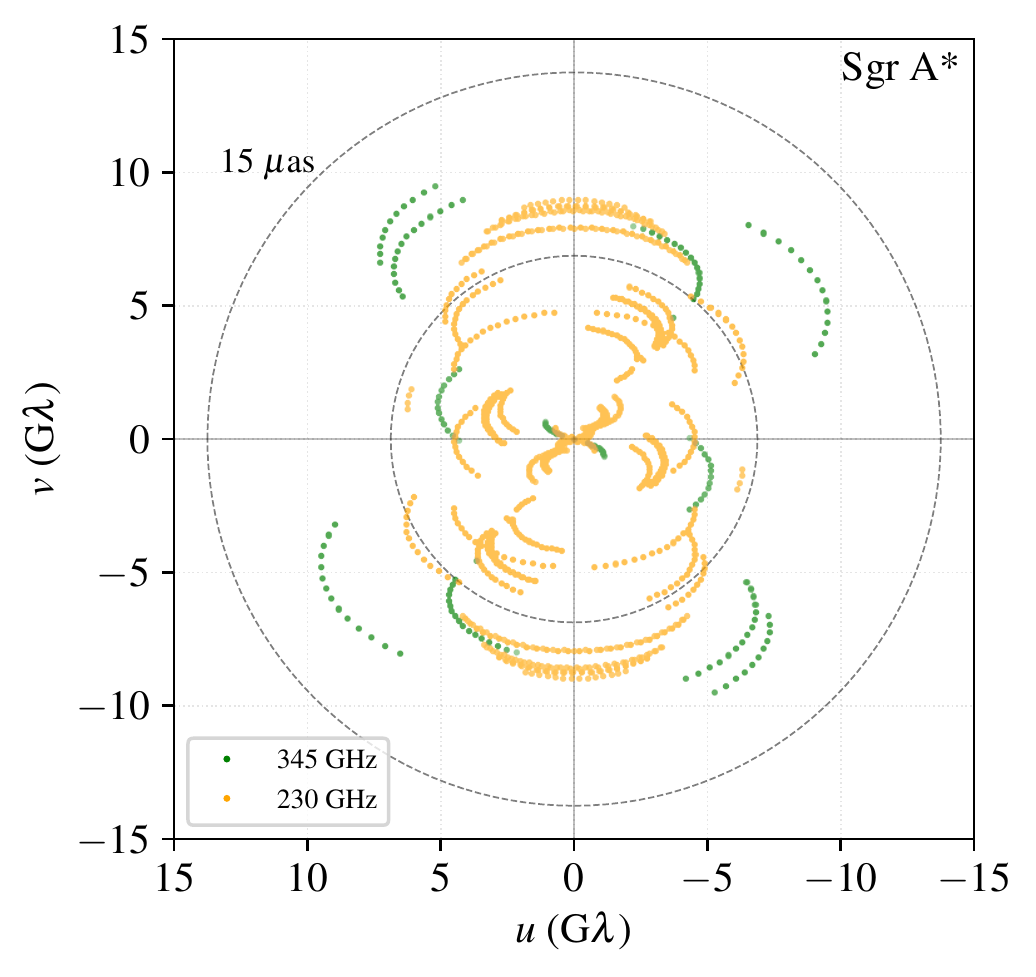}
    \caption{(Top) Current EHT array (2023).  (Bottom) Interferometric coverage for \m87 and \sgra at 230 \& 345\,GHz, assuming April observing conditions.  The coverage reflects estimated detections made through simulating \m87 and \sgra models at both frequencies with the EHT array as fielded in 2023 (see \autoref{tab:Array}, and \autoref{sec:SyntheticData}).  {\it Note that for the EHT in 2023, 230\,GHz and 345\,GHz observations cannot be made simultaneously, so the coverage shown cannot be combined to form a full image (as is possible in the ngEHT Phase 1 and Phase 2 arrays)}. The opacity of each plotted data point is proportional to how frequently it is expected to be detected.  The outer and inner dashed circles mark baseline lengths corresponding to angular scales of 15\,\uas and 30\,\uas, respectively.}
    \label{fig:eht2022_globe_uv}
\end{figure*}

\begin{figure*}[htbp!]
    \centering
    \includegraphics[width=1.00\textwidth]{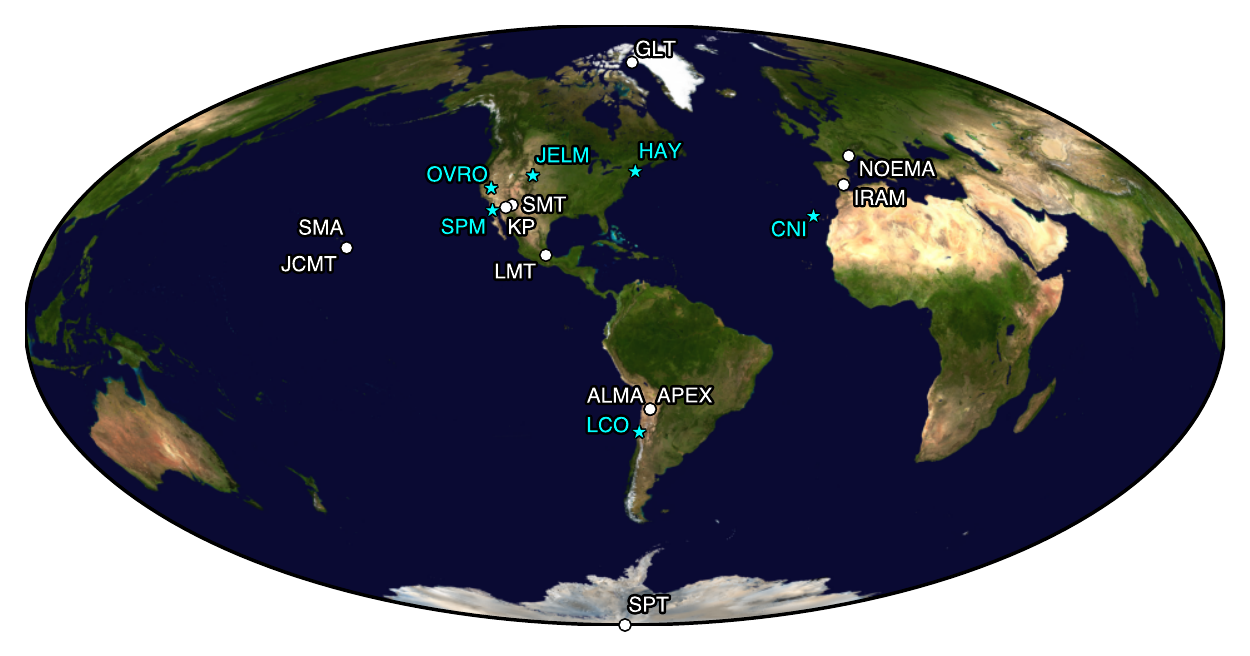}
    \includegraphics[width=0.49\textwidth]{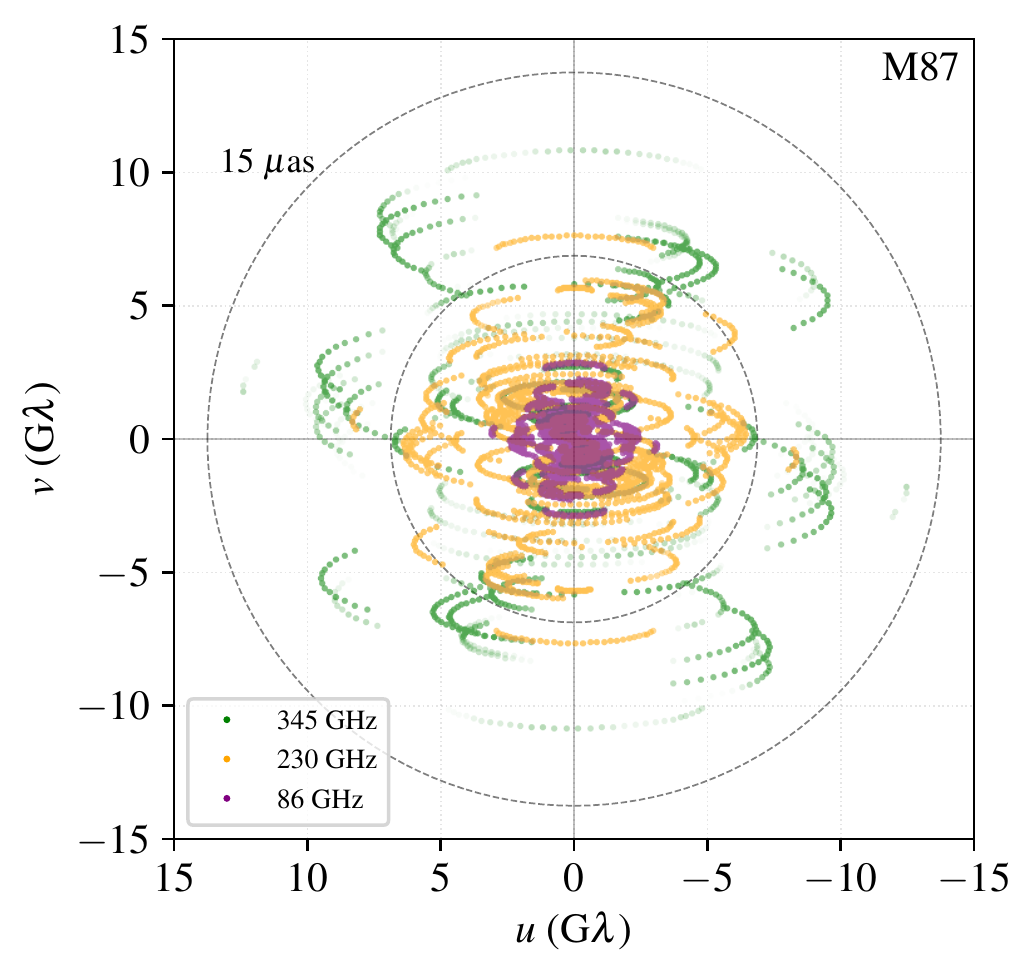}
    \includegraphics[width=0.49\textwidth]{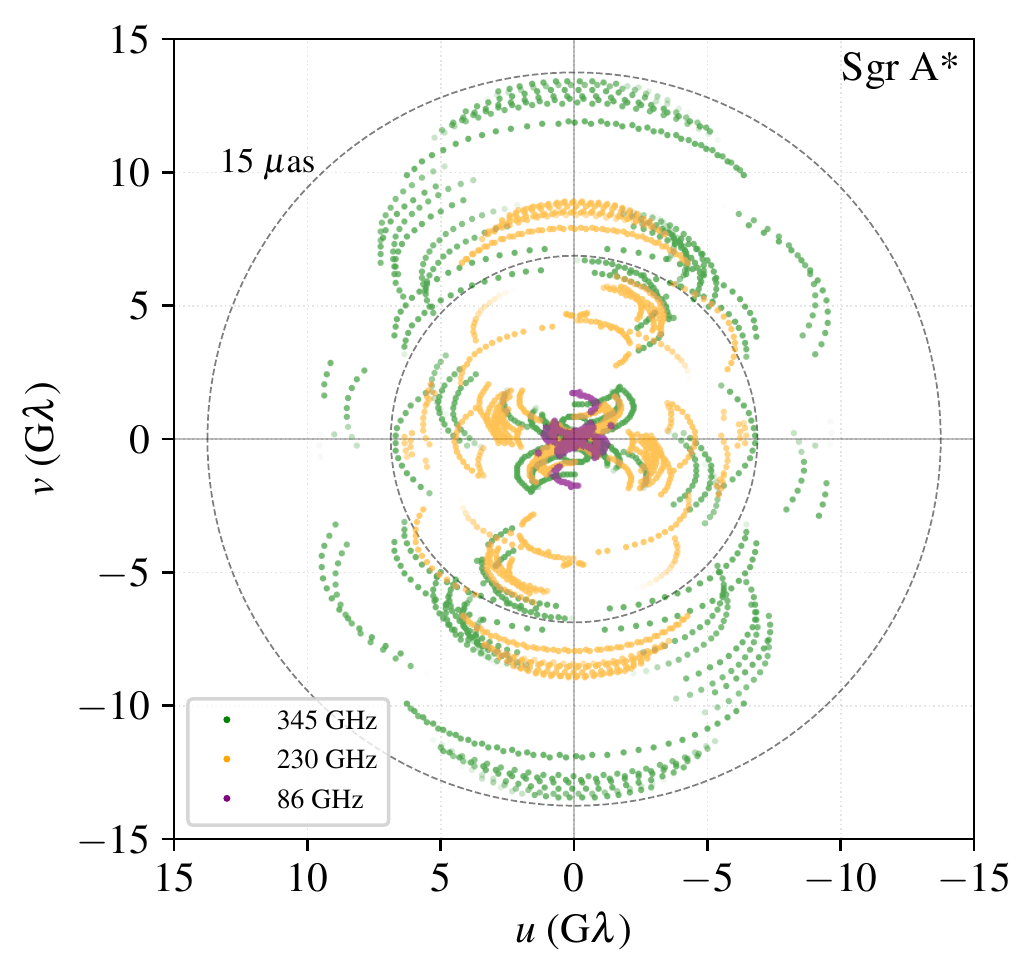}
    \caption{(Top) ngEHT \phaseone array; white sites are current EHT dishes, blue sites are ngEHT sites.  (Bottom) Interferometric coverage for \m87 and \sgra at 86\,GHz, 230\,GHz, and 345\,GHz, assuming April observing conditions.  The coverage reflects estimated detections made through simulating \m87 and \sgra models at all three frequencies with the ngEHT Phase 1 array (see \autoref{tab:Array} and \autoref{sec:SyntheticData}).  Sites without multi-frequency capabilities are assumed to be observing only at their highest frequency.  The opacity of each plotted data point is proportional to how frequently it is expected to be detected.  The outer and inner dashed circles mark baseline lengths corresponding to angular scales of 15\,\uas and 30\,\uas, respectively.}
    \label{fig:ngeht_phase1_globe_uv}
\end{figure*}

\begin{figure*}[htbp!]
    \centering
    \includegraphics[width=1.00\textwidth]{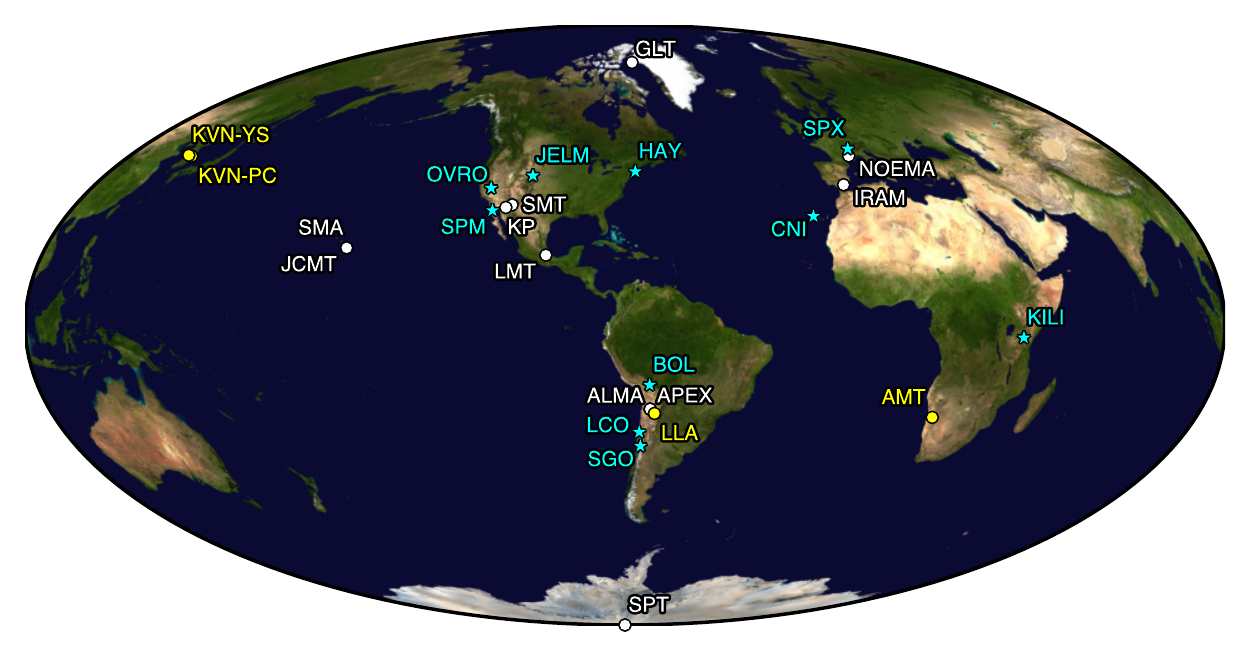}
    \includegraphics[width=0.49\textwidth]{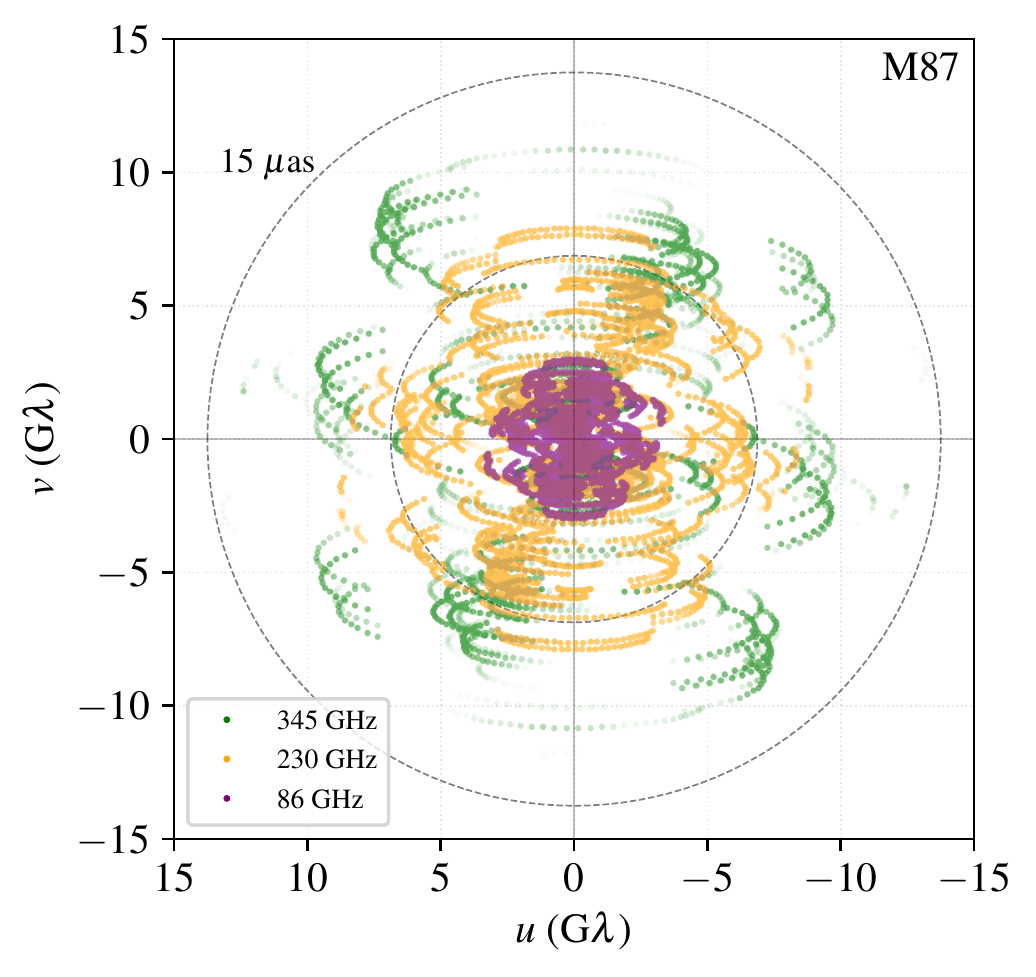}
    \includegraphics[width=0.49\textwidth]{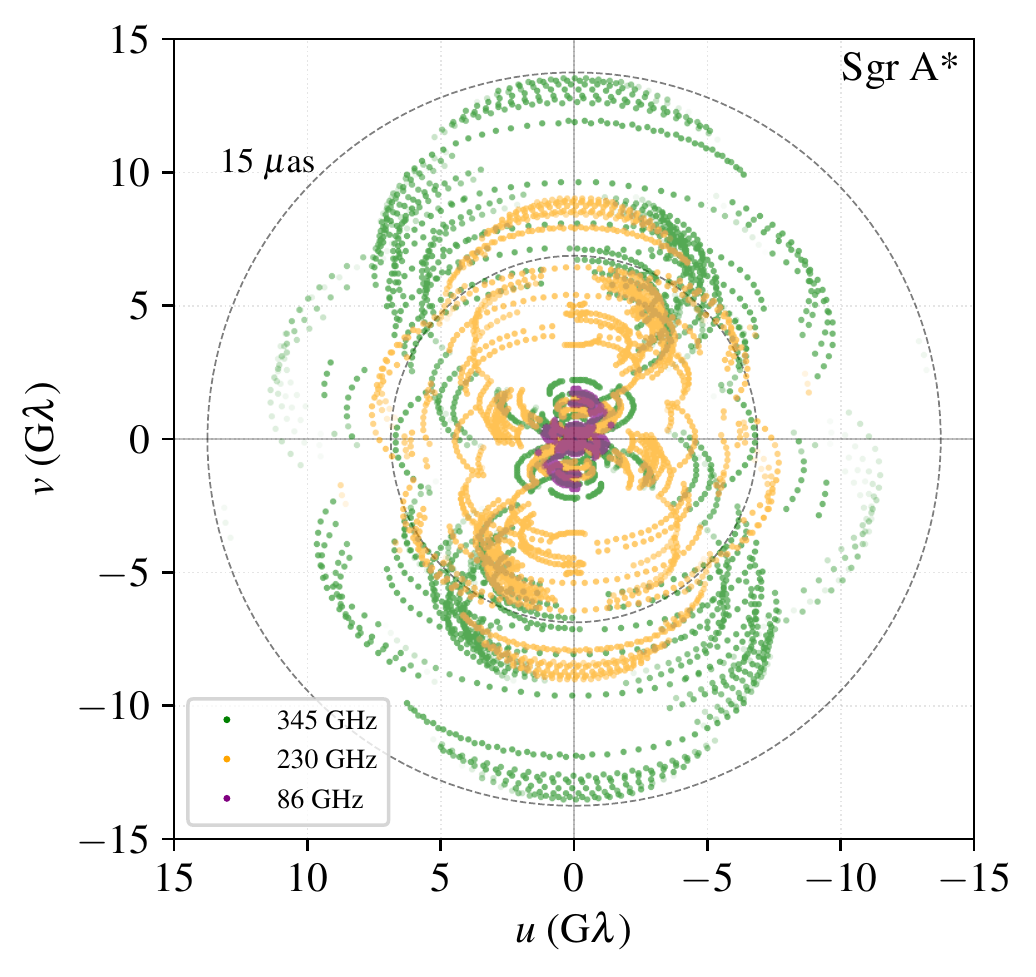}
    \caption{(Top) ngEHT \phasetwo array; white sites are current EHT dishes, blue sites are ngEHT sites, and yellow sites are planned or existing facilities that may join (ng)EHT observations.  (Bottom) Interferometric coverage for \m87 and \sgra at 86\,GHz, 230\,GHz, and 345\,GHz, assuming April observing conditions.  The coverage reflects estimated detections made through simulating \m87 and \sgra models at all three frequencies with the ngEHT Phase 2 array (see \autoref{tab:Array} and \autoref{sec:SyntheticData}). Sites without multi-frequency capabilities are assumed to be observing only at their highest frequency.  The opacity of each plotted data point is proportional to how frequently it is expected to be detected.  The outer and inner dashed circles mark baseline lengths corresponding to angular scales of 15\,\uas and 30\,\uas, respectively.}
    \label{fig:ngeht_phase2_globe_uv}
\end{figure*}

Simulated EHT coverage for the array fielded during the 2023 observing campaign is shown in Figure~\ref{fig:eht2022_globe_uv}.  The enhanced baseline coverage that will be provided by the ngEHT Phases 1 and 2 is shown in Figures~\ref{fig:ngeht_phase1_globe_uv} and \ref{fig:ngeht_phase2_globe_uv}.  In Figure~\ref{fig:ngeht_phase1_globe_uv}, the JELM station has been added to reflect a Phase 1 array as described in \autoref{sec:AlternateStaging}, while Figure~\ref{fig:ngeht_phase2_globe_uv} shows the array as described in \autoref{sec:SiteSelectionPhase2}.  

\section{Operating Modes}

Key Science Goals (KSGs) motivate five basic operation modes of the ngEHT, which enable specific science use cases.  Details and constraints of each mode are defined by cost/benefit analyses and feasibility studies. Factors to be considered in this analysis include time allocation at various sites, weather, data throughput with implications for disk inventory and correlation, reliability and up-time, and maintenance strategy.   The following five subsections provide a narrative summary for each of the five envisaged operating modes of the ngEHT, which are then summarized with salient characteristics in table \ref{tab:OpsModes}.

\begin{center}
\begin{table*}
\begin{tabular}{||c c c c||} 

 \hline \hline
 OpsMode & stations in array & cadence \& duration & science case \\ [0.5ex] 
 
 \hline\hline
 Campaign & 14 to 21 & 1x7  & Sgr~A*, M87, blazars, jets \\ 

 \hline
 Long term  & 5 to 20 & 5 days,  & \m87 \& blazar kinematics, \\
 monitoring &  & 3 to 7 mo  & \sgra flares \\
 \hline
 Target of & 3 to 6  & 1 wk & flares, \\
 Opportunity & & 3x/yr & gravitational waves \\
 \hline
 CMF & 14 to 21 & ~hours x 2 wks & AGNs, black hole binaries \\
 \hline
 Beyond  & 1 to 10 & dependent on & stellar birth,\\ 
 ngEHT &  & science case & fast radio bursts\\
 \hline\hline
 
\end{tabular}
    \caption{The five ngEHT operating modes and selected salient characteristics of each.}
    \label{tab:OpsModes}
    \end{table*}
\end{center}

\subsection{Campaign}
This is a single epoch annual multi-day campaign, which is an extension of the standard annual campaign already executed by the Event Horizon Telescope (EHT). The 11 EHT sites defined by those used in the 2022 EHT array are assumed to participate with the ngEHT sites.   In this mode, dedicated tracks are based on clearly defined, community-prioritized science cases, in some cases led by a principal investigator.  

The campaign mode pursues M87 and Sgr~A* science cases with enhanced capability relative to EHT due to improved sensitivity and great uv coverage from the larger 21 site array.  This results in enhanced M87 imaging, snapshot sensitivity for Sgr~A* movies, and studies of blazar jet collimation.

\subsection{Long Term Monitoring}
The long term monitoring mode uses extended duration and more frequent cadence observations with a smaller subset of the existing EHT sites participating. The ngEHT sites enable this mode through their purpose-designed flexibility and dedicated time allocation for VLBI.

Several multi-week observations over the course of the year once again have dedicated tracks based on clearly defined, community-prioritized science cases. These science cases are in the broad areas of \m87 movies, blazar kinematic studies, and Sgr~A* flaring activity monitoring. As an example, to continuously track changes in the \m87 appearance (\m87 movies), reconstructing images separated by the expected coherence timescale ($\sim 50 GM/c^3 \approx$ 20 days) is needed. A single-year EHT campaign may only last about a week \citep{Paper1} -- too short for a significant change in the source appearance, while combining results from separate years only provides uncorrelated source snapshots, without the ability to track continuous motion of the flow features \citep{Wielgus2020}. Similarly, in the published EHT analyses of blazar observations \citep{Kim_2020,Issaoun_2022, Jorstad_2023} short duration of the EHT campaigns, and the lack of repeated observations on timescales of weeks or months, has been recognized as the main factor limiting the current EHT ability to study jet kinematics.

\subsection{Target of Opportunity}
Target of Opportunity (ToO) is an agile operational follow-up by ngEHT to an unpredictable event observed with another facility. It involves ad-hoc subarrays of the 11 existing EHT sites - those which are available - while all of the ngEHT dedicated sites will be made available for suitably scientifically interesting ToO observations. Broad science areas are expected to be in the area of  flares, gravitational waves, and fast radio burst counterparts.

\subsection{Coordinated Multi-Facility}
The Coordinated Multi-Facility (CMF) mode is characterized by coordinated, multi-facility, multi-messenger observations involving multiple ngEHT sites and at least one other ground or space instrument (e.g., Chandra, the GRAVITY instrument, and any of various optical/IR facilities).  This CMF mode is a planned continuation of the successful EHT Multi-Wavelength campaigns \citep[see ][]{M87_MWL_2021}.

The broad science areas are expected to be multi-wavelength studies of Active Galactic Nuclei, binary and  singular black holes.

\subsection{Beyond-ngEHT}
This single dish mode covers any observation that is performed outside the core ngEHT science mission, but will still be part of the ngEHT operating model due to local institutional requirements or synergies with other communities or facilities.   

Science is expected to be in the broad area of star forming regions, fast radio bursts, and astronomical maser studies of transitions in the ngEHT RF bands.
\section{Data Processing}

The next-generation (ngEHT) expands upon the existing 11-station EHT with around 10 additional small-dish antennas as well as simultaneous 230/345 GHz observations. In addition to the roughly $\sim$10-fold increase in aggregate data rate across the entire array, the ngEHT is expected to operate as a full-season agile observatory as opposed to the $\sim$few observing days per year of the current EHT. When all participating sites are observing, one night of ngEHT produces around $\sim$10 PB of raw data (around 0.5 PB per site), resulting in up to a $\sim$couple hundred PBs per year that must be processed. An efficient streamlined approach to data processing and management is required to facilitate media turn-over and to deliver quality assured science-ready data products in a timely manner.

\begin{figure}
    \centering
    \includegraphics[width=1.0\columnwidth]{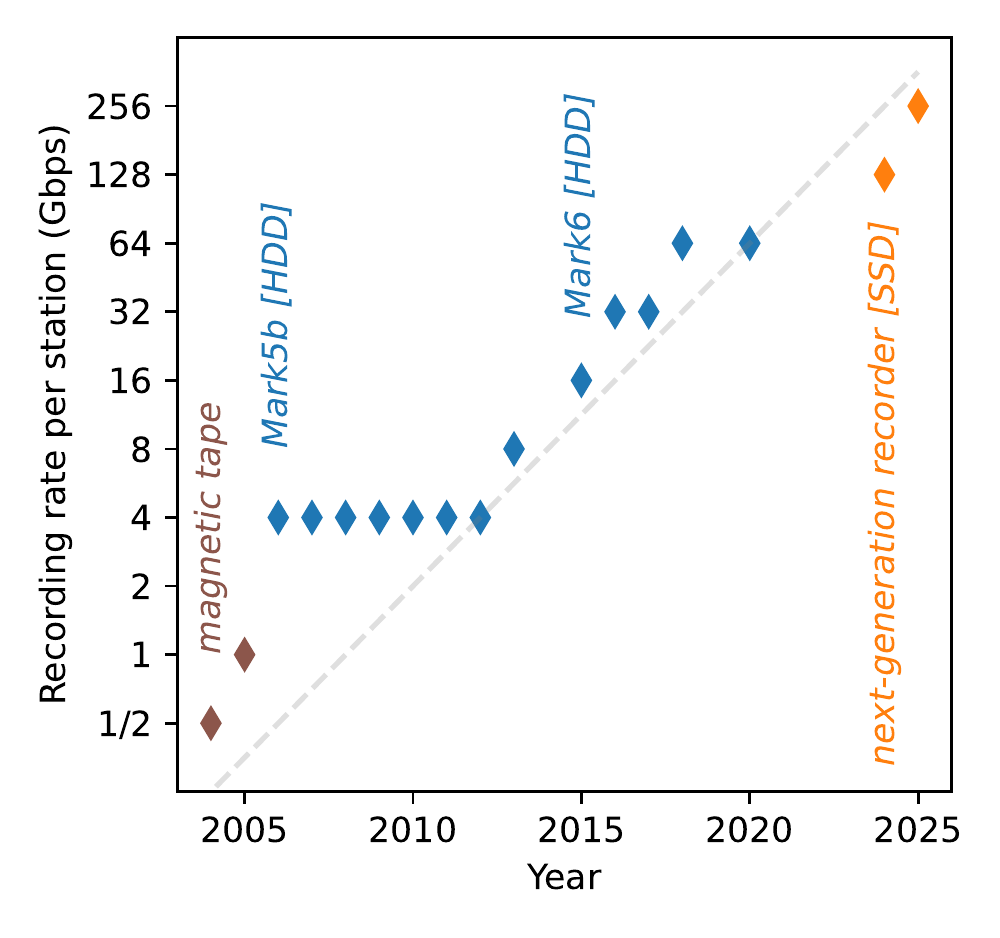}
    \caption{EHT/ngEHT data rate per station over two decades, roughly doubling every two years. The large bandwidths provide the EHT/ngEHT the necessary continuum sensitivity for ultra-high resolution VLBI imaging at \mbox{(sub-)1\,mm} using a highly heterogeneous network of telescopes. Maintaining this trend has required the regular adoption of commercial technologies as they became available.
    }
    \label{fig:eht_bandwidth}
\end{figure}

The large data rates and volumes of the ngEHT motivate continued adoption and assimilation of new technologies, which has allowed a rough tracking of Moore’s law over two decades of global mm-VLBI development (Figure~\ref{fig:eht_bandwidth}). On the timescale of a $\sim$decade, we anticipate a transition from Hard Disk Drives (HDDs) to Solid State Disks (SSDs) for recording and eventually transport, which provides high-bandwidth, high-density, and power-efficient I/O. SSDs carry a gradually narrowing cost premium of 5--10x versus HDDs in \$/TB, but their use would allow ngEHT recording systems to keep up with the ngEHT data rates while staying within practical power, weight, and space footprints for efficient media handling, staging, and transport.

GPU’s have become the platform of choice for massively parallel vector/tensor calculations due to their efficiency and ease of use, and they are being researched or already adopted for efficient VLBI correlation across several experiments. The ``embarrassingly parallel" nature of VLBI correlation is suitable for high-throughput computing (HTC) workflows, and the irregular scheduling of VLBI observations means that on-demand scalable computational resources are desirable.

\subsection{Data Transport}

While observing, the ngEHT will produce an aggregate $\sim$5 Tbps of digital signal data that must ultimately be transported from remote sites to a central location for processing. Similar to the EHT, the only currently available means for moving such a large total volume of data from the (sometimes very-) remote locations in a reasonable amount of time is by physical transport of recorded media. VLBI experiments such as the European VLBI Network\footnote{\url{https://www.evlbi.org/}} (EVN) are currently able to transport data electronically, due to considerably lower data rates and more accessible sites (typically at sea level) that are linked to a high speed internet backbone. The ngVLA reference design \citep{ngvla} also includes real-time data transport (320 Gbps per antenna) and correlation via ground fiber (both dedicated and leased commercial), even for the longest baselines spanning the United States and territories. However because the ngEHT operates a (comparatively) small number of antennas at remote locations spanning the globe, shipment of physical media is expected to remain the fastest and most economical method of transferring 100s PB of data for the foreseeable future. Consistent array-wide high-speed internet access, such as that provided by global commercial Satellite RF internet, will nevertheless be extremely useful for rapid transfer of small amounts ($\sim$1\%) of data for interferometric validation and for obtaining near-realtime results where scientifically relevant.

The ngEHT is designed to operate full-season, and this motivates a rapid processing and recycling of recording media to limit costs. Media are expected to be redeployed approximately once per two months (on average), versus once per 2-3 years as for the current EHT. As a result, there is less of a focus on media utility for economical long-term storage, and more toward efficient recording and transport. Once data are brought to the correlation facility, they can be offloaded to local HDD-based storage if needed, for example in the case of experiments including the South Pole Telescope which can incur several months of shipping delay. A rotating media library of 200 PB would be required to support bimonthly turn-around of observations totaling 10 PB every three days while providing ample time for average shipping time and data offload.

\subsection{Correlation}

Correlation is the process of calculating pairwise correlation coefficients between the signals captured at each antenna. Because this is an operation on the PB of raw VLBI data, it is both I/O and computationally intensive and requires carefully matched computing platforms for effective processing. Correlation coefficients are typically calculated in the frequency domain using a so-called FX correlation architecture that enables efficient searching over unknown time delay via Fourier convolution. Frequency domain processing also allows for convenient matching of signals from partially overlapping bandwidths as well as the application of linear and non-linear corrections to align the data. The consequences of an FX architecture is a large up-front cost to data channelization, scaling linearly with the number of antennas. For a 20-station network at ngEHT bandwidths, the $O(N)$ cost from data stream pre-processing and the $O(N^2)$ cost from calculating all pair-wise correlations are expected to be roughly comparable.

The current EHT records at 64 Gbps over 11 stations, for an aggregate rate of 0.7 Tbps. Data are correlated at dedicated computing clusters at MIT Haystack Observatory and the Max Planck Institute for Radioastronomy using the DiFX software correlator \citep{deller2011}. In aggregate, $\sim$2.5k cores are able to process the full EHT bandwidth at about 10\% real-time. Scaling linearly to the aggregate data rate of ngEHT requires $\sim$20k cores to process ngEHT's $\sim$5 Tbps at 10\% real-time (in comparison, 300 hours of data per year is a reasonable upper limit for ngEHT data throughput and reflects a duty-cycle of ~3.5\%). A quadratic scaling with the number of stations would imply double the requirement, but this can be balanced against $\sim$5-10\% year-over-year improvements to single-core performance. CPU core density and efficiency are also increasing at a much faster rate, and GPU acceleration of both channelization and cross-multiply stages of correlation are expected to increase efficiency by another factor of $\sim$several. A detailed description and modeling of VLBI software correlation performance is presented in \citet{2022PASP..134j4501V} alongside several benchmark results including those from the literature.

Approximately $\sim$60M CPU core-hours would be required to correlate $\sim$680 PB (300 hours) of raw data. VLBI data are taken non-continuously throughout the year and sometimes require multiple passes through correlation to iterate on a proper configuration. Thus is it necessary to over-provision on-demand computational resources by a factor of $\sim$few in order to avoid backlogs and ensure regular turn-around of recording media. Around $\sim$100k on-demand CPU cores would be appropriate to keep up with the largest projected ngEHT data volumes, which is the size of a large institutional research cluster or a few medium-sized clusters distributed geographically. Due to the over-provisioning, the resources are ideally time-shared with other computing requirements (calibration and imaging, other VLBI correlation, or other general uses).

\subsection{Calibration and Reduction}

Output from correlation is at a resolution of $\sim$1\,MHz in bandwidth and $\sim$1\,second in time, which is required to capture residual instrumental and environmental systematics that affect the measured correlation coefficients such as lines, frequency response, relative delays, and time-varying gains and atmospheric phase \citep{Paper2,Paper3}. These products are smaller than the recorded VLBI signals by a factor of ~$>$10$^3$ due to the large amount of accumulation following cross-correlation. A calibration process then solves for a refined instrument model and folds in any additional priors on the instrument response.

A key element of the calibration process is ``fringe-fitting'' where a parameterized phase model (typically relative {\em delay} and {\em delay-rate} over a short time interval) is self-calibrated to the correlator output. The fitting process verifies that a correlated signal exists in the data, measures the correlation coefficient, and allows data to be further coherently averaged, reducing the overall data volume by another factor of $\sim$10$^4$. Dedicated fringe fitting and calibration pipelines \citep{ehthops,rpicard} were developed for EHT data to address the heterogeneous nature of the array and unique data properties. Compared to correlation, the computing requirements to fit a basic phase calibration model are low. For example the EHT 2017 campaign data set (5 nights, 8 stations) can be processed through a multi-stage calibration and reduction pipeline using $\sim$1.5k CPU core-hours \citep{ehthops}.

This initial stage of calibration and reduction is aimed at reducing the overall data volume and complexity for downstream data products, while applying only well-determined calibration solutions. Since data are manipulated and averaged, it is important to avoid introducing calibration solution noise or detailed assumptions about the source. In cases where calibration solutions are under-determined or degenerate with source model parameters, they must be jointly modeled during analysis. The complexity and computational cost can increase dramatically due to the high dimensionality of an instrument model, particularly for the case of formal Bayesian inference \citep{themis,dmc}.

\section{Instrumentation Design}
\begin{figure*}
    \includegraphics[width=1\textwidth]{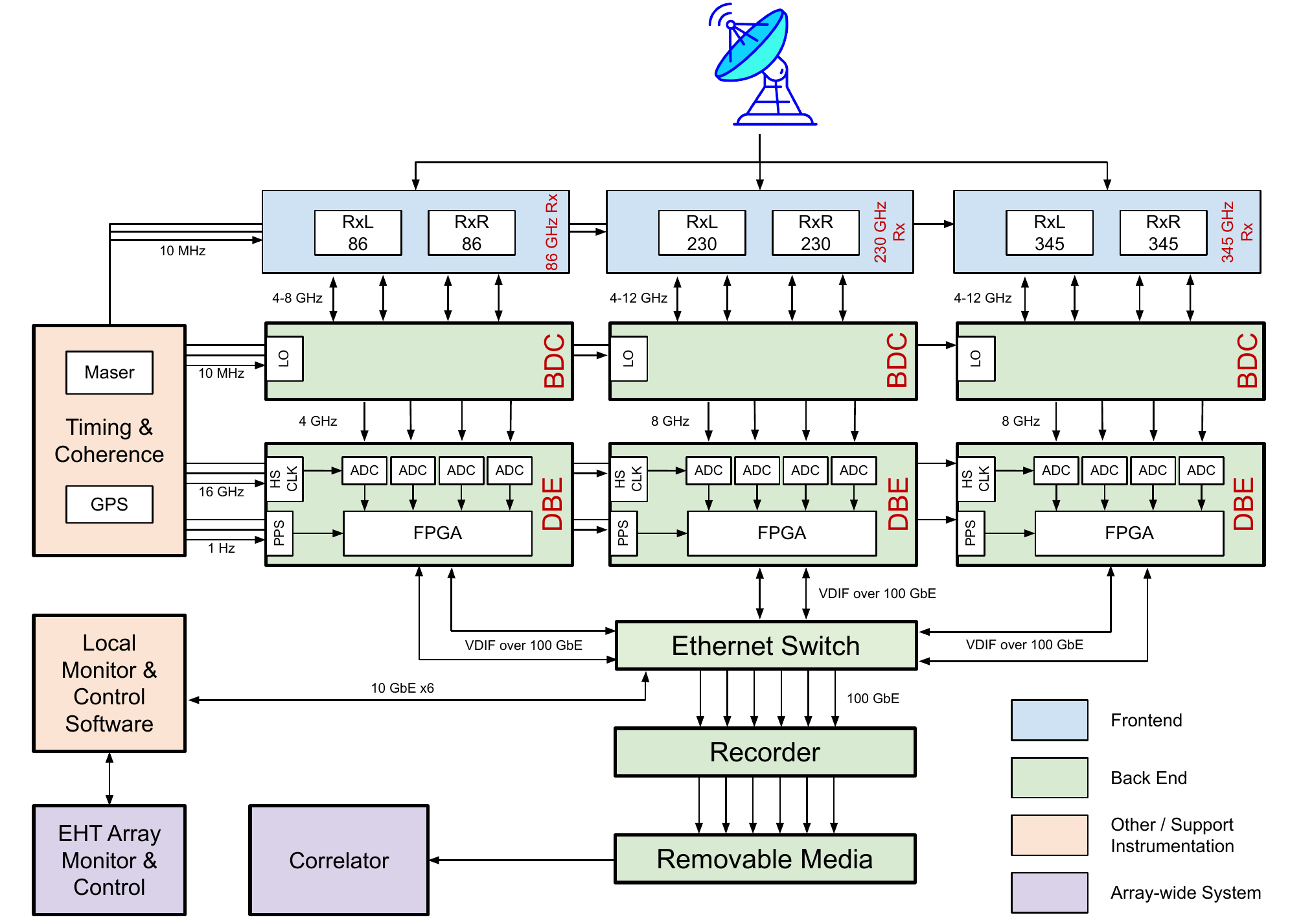}
    \caption{ 
    Functional block diagram of a next-generation EHT station. All elements shown in the figure are either commercially available (e.g., Hydrogen Maser), or in advanced prototyping stages, and suitable for deployment at ngEHT stations.  The Timing \& Coherence block consists of a Maser and GPS system, which provides ultra-stable clock signals for the DBE and references for the dual-polarization receivers and the BDC.  A high-speed ethernet switch routes DBE packets to recorders with modular/removable media for shipment to the central correlator for interferometric processing.  ngEHT Monitor and Control is handled by local and global systems. 
    }
    \label{fig:ngEHT_station_blk_diag}
\end{figure*}

In this section we describe the basic elements of the ngEHT system (see Figure~\ref{fig:ngEHT_station_blk_diag}.  These are the result of several internal project reviews, including a Systems Requirements Review, held on June 9-10, 2022.  At this stage of the project, the ngEHT team has developed initial instrumental requirements through a process of preliminary trade-off analysis.  This process has enabled development of several prototypes, which have been selected for deployment in \phaseone of the project, and these specific elements of the ngEHT system are described below.

\subsection{Receiver} \label{sec:Receiver}

\begin{figure*}
    \includegraphics[width=0.57\textwidth]{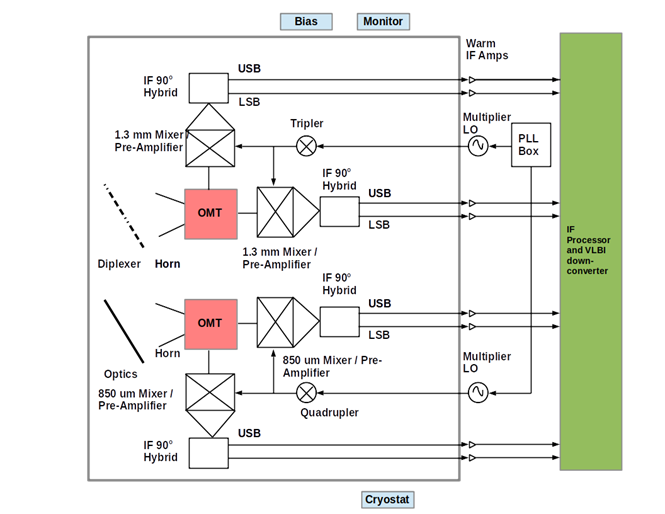} \includegraphics[width=0.43\textwidth]{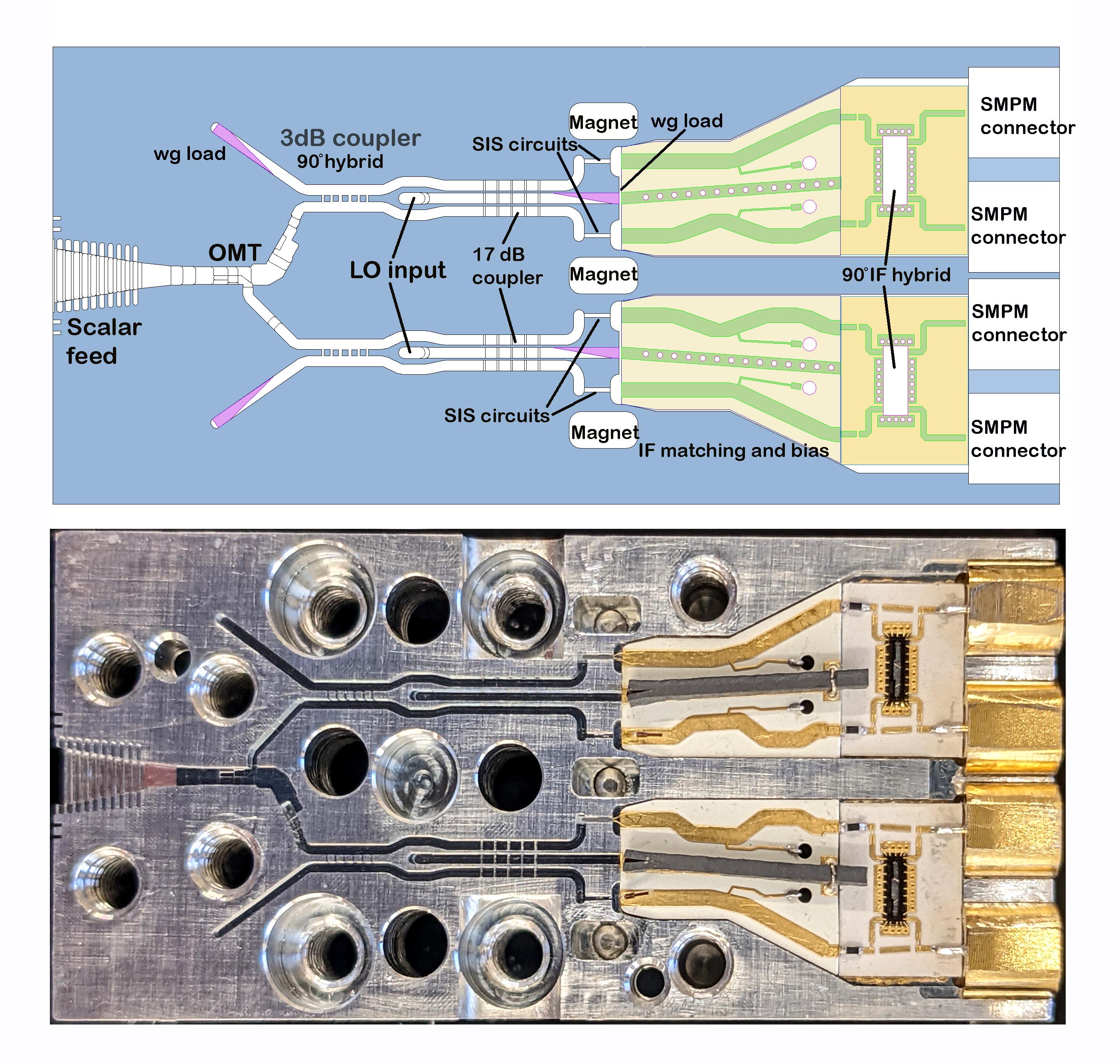}
    \caption{ 
    {\bf Left:} Block Diagram of the proposed dual band SIS receivers. Both the 1.3mm and 850 µm band receivers will be built inside a single cryostat. \\
    {\bf Right Top:} Schematic outline of the the 1.3mm frontend receiver block. This block shows the cold section of the corrugated square feed-horn feeding an orthomode transducer (OMT) section that separates the input signal into two polarization channels, one in each of the top and bottom halves of the block. In each polarization, there is a RF $90^\circ$  hybrid followed by LO couplers, ending in two SIS junctions. The IF outputs of the pair of SIS junctions passes through IF matching and bias tee to a superconducting IF $90^\circ$ hybrid, which outputs the upper and lower sideband IF signals from that channel. In all 4 SIS junctions are used in each mixer block, with Cooper pair tunneling suppresesed by permanent magnets.\\
    {\bf Right Bottom:} Photo of one half of an assembled 1.3 mm fronted receiver block. 
    }
    \label{fig:dualband_schematic}
\end{figure*}

In Figure~\ref{fig:dualband_schematic} (left), we present a block diagram of a dual frequency receiver being constructed for ngEHT and to be deployed at the LMT. A single cryostat will hold two different receivers and the two different frequency bands are sent to each receiver through a frequency diplexer. Each receiver is dual-polarized, and features sideband separation mixers (see Table~\ref{table:dual_frequency_rx_spec}). Both bands illuminate a single beam on sky, and the overall dual-frequency receiver has eight IF outputs, each of which is 4--12 GHz wide.

\begin{table}[htbp!] \centering
\begin{tabular}{ll}
\hline
{Item} & {Description} \\
\hline
3mm RF Band & 82 - 116 GHz\\
1mm RF Band & 210 - 280 GHz\\
850$\mu$m RF Band & 275 -- 375 GHz\\
Polarizations & Dual pol in each band\\
Sidebands & 2SB Receivers in each band\\
IF Frequency & 4 -- 12 GHz (1mm,0.85$\mu$m)\\
&4 -- 8 GHz (3mm)\\
Receiver Noise & $<$ 50 K in 3mm band\\
\ \ Temperature & 60 -- 70 K in 1mm band\\
&70 -- 80 K in 850$\mu$m band\\
\hline
\end{tabular}
\caption{Specifications of the ngEHT multi-band Frequency Receiver}
\label{table:dual_frequency_rx_spec}
\end{table}

In an effort to make the design highly modular and scalable to reproduce for additional new telescopes of the ngEHT array, considerable effort has been invested into making the mixer block compact and highly integrated. In Figure~\ref{fig:dualband_schematic} (right), we show the components of this highly integrated block. Shown is a photo of the bottom block of a split-block mixer (bottom) and a schematic diagram of the components (top). A similar design will be employed for the 850 $\mu$m receiver as well. The 4 IF outputs from each of the mixer blocks are amplified cryogenically using commercially available low-noise amplifiers. 

Each of the receiver bands are equipped with independent local-oscillator (LO) systems. YIG oscillators are lower frequencies (in the 18-30 GHz) range are multiplied up to the 3mm wavelength band, and subsequently amplified using W-band power amplifiers. This is then fed through cryogenic triplers to produce the required LO signal. The drain currents of the last stage of the W-band power amplifiers can be adjusted to set the appropriate LO power for the mixers. The whole LO system is phase locked, and fully computer controlled with no mechanical moving parts.

Implementation of additional 86GHz capability to enable Frequency Phase Transfer (FPT) will proceed along multiple paths.  For existing sites that already field 86 GHz receivers, these will be coupled where possible to higher frequency receivers using dichroics that enable simultaneous operation (e.g., GLT, JCMT).  At existing sites that do not have 86GHz receivers, or where existing 86~GHz systems cannot be used, new HEMT-based 86~GHz receivers, cooled to 20K, will be added and coupled via dichroics.  These new 86GHz receivers will follow existing and proven designs.  Finally, for the new ngEHT sites, a tri-band dewar that incorporates 86, 230 and 345 GHz receivers will be constructed, following existing designs and prototypes for the ongoing upgrade of the Submillimeter Array in Hawaii.
  
\subsection{Backend }
The ngEHT backend consisting of the Block Down Converter (BDC) and the Digital BackEnd (DBE) will process four times the instantaneous bandwidth (dual sideband, dual polarization, and dual frequency) of the current EHT.

The BDC  performs a frequency translation and signal conditioning of the analog signal from the receivers. The Intermediate Frequency (IF) signal is converted to baseband, and output power levels are adjusted to optimally load the Analog to Digital Converter (ADC). The design of this BDC  was initiated  and  functionality was implemented in a prototype, constructed by Xmicrowave LLC. The prototype was manufactured using drop-in PCB (Printed Circuit Board) modules instead of connectorized components, which is more representative of the final BDC PCB.  A full characterization has been conducted and the results meet the required specifications. The final BDC will consist of integrated PCB  units instead of discrete drop-ins.

The DBE prototype currently being used for testing and development is a two board system. This prototype uses a custom circuit board holding four ADCs, which digitize the analog signal from the BDC. This board sends the digital data stream to a commercial evaluation board,  the VCU128 which houses the VU37P Field Programmable Gate Array (FPGA) from the Virtex Ultrascale+ family manufactured by Xilinx.  Each 4 bit ADC is clocked at 16.384 GHz. The Nyquist bandwidth of this system is therefore 8.192 GHz, which is interoperable with the current EHT. The evaluation board is useful for current tests and development, and it will be replaced with a custom board design; the design of this new board is underway with an estimated one-year timeline to completion. Parts are being acquired to support a build of five units.

In addition to hardware (board) development, the initial firmware command set  has been successfully completed, including an ADC interface  module, a requantization block from 4 bits to 2 bits in the processing module, a packetization module, a 100 Gb transmission module, a Universal Asynchronous Receiver Transmitter(UART) monitor and control module, and a timing module. Further features that will be included in the firmware are channelization, 1 Gb monitor and control, and slope and ripple equalization. 

With 2-bit quantization and Nyquist sampling, each DBE can process the full IF bandwidth (8 GHz) from the 1mm and 0.85$\mu$m band receivers for a total data throughput of 128 Gb/s.  For the 3mm band, a narrower IF bandwidth (4 GHz) is sufficient to achieve Key Science Goals and Frequency Phase Transfer calibration.  At 3mm, the resulting data throughput is 64 Gb/s.

\subsection{Recorder }

The recorder is expected to be based around a set of commercial off-the-shelf (COTS) components hosted on a commodity multi-processor
computer running a GNU/Linux operating system with a PCIe 4.0 interconnect. A single recording unit is matched to one or more streams from the digital back-end system (DBE), which is designed to output 64 Gbps data streams on 100 GbE interconnect using the VDIF transport protocol \citep[VTP,][]{vtp} over UDP. Specialized software on the recording unit provides efficient network capture at the required rates, simple packet inspection to ensure data continuity and integrity, distributed writing of VDIF file streams to disk, and an interface to the VLBI monitor and control system.

The host recording system will buffer the incoming data in system RAM, while simultaneously draining this data to persistent memory for storage. The persistent storage is expected to be a set of solid-state drives (SSDs) attached via PCIe/NVMe (integrated media). The total number, individual capacity and write performance of the component SSDs in the persistent memory pool will be selected such that they are sufficient to absorb the total aggregate data rate and meet the desired overall capacity and cost constraints. In order to facilitate playback of detachable data modules for subsequent correlation or transfer, the recorder will maintain a file system on the media so that may be mounted by separate machine.

Although an SSD-based recorder has several advantages over a HDD-based system in terms of speed, power, density, weight, and latency, SSDs are anticipated to carry a significant cost premium to HDDs for the next decade. Moreover a modular removable disk pack system analogous to the semi-custom Mark6 module \citep{Whitney2013} has yet to be designed, which limits the flexibility of current COTS SSD recorders. For this reason, large volume data storage and possibly transport may still rely on HDD-based solutions for some time, with SSD-to-HDD data offload capability at site or at the correlator.

\begin{table}
\centering
\begin{tabular}{ccc}
\hline
 &  Mark6 & ngRecorder \\
\hline
rack space & 11U & 2U \\
disks & 32 HDD & 24 SSD \\
capacity & 512 TB & 369 TB \\
interface & 4x 10/25 GbE & 2x 100 GbE \\
rate & 16/32 Gbps & 128 Gbps \\
hours at rate & 71.1/35.6 & 6.4 \\
disk modules & yes & no \\
\hline
\end{tabular}
\caption{Specifications for a modular VLBI recorder, including those of the Mark6 \citep{Whitney2013} currently in use across the EHT. Reference specifications for a next-generation SSD-based recorder are based on common currently available COTS SSD storage servers.}
\end{table}

\subsection{Array Monitoring and Control} \label{subsec:ArrayMC}
The operations concept for the ngEHT extends beyond the single annual campaign of the current EHT:
\begin{itemize}
 \item 60 nights of observing per year
 \item Up to 21 stations observing simultaneously
 \item Varied observation cadences and durations throughout the year
 \item Readiness for VLBI observing in 24h or less to capture ToOs
 \item Multi-messenger campaigns
 \item Configurable subarrays
 \item As much remote operation as possible
\end{itemize}

This model and its increase in capability has direct impact on the requirements and subsequent complexity of the M\&C system for the ngEHT. As the M\&C system serves as a main user interface point for the array, its design must be user-centered and have due consideration for human factors concerns. As well, the operations concept is designed to address an explicit need, voiced at the ngEHT Operations Workshop (31 Mar 2022), to reduce the burden (relative to 2022 EHT operations) for on-site monitoring, control, and maintenance of VLBI equipment. The areas to address include differing methods of monitoring and control for each station and heavy reliance on local operations at each site, including the need for VLBI specialists on site.

As the first ngEHT sites are brought online, they will participate in the annual EHT observing campaign. To facilitate this participation, the M\&C system will be compatible with the EHT operations plans and procedures by relaying data to the existing VLBI Monitor server, providing remote control of station subsystems, and providing status, logs, and metadata as required. Outside of the annual EHT campaign, the ngEHT operations concept calls for an annual monitoring campaign where the M\&C system will be used to operate and monitor the entire array. It will provide a uniform and cohesive monitoring and control experience to the array operators while managing a heterogeneous array of ngEHT stations and stations that use the existing EHT VM\&C system and backend equipment.

Collecting observation metadata from a heterogeneous array of telescopes that have non-standardized interfaces for M\&C and data collection is a significant design and operational challenge. To take advantage of the opportunity presented by the ngEHT designing its own telescopes, it is expected that the M\&C component of the telescopes for ngEHT sites will be designed in conjunction with the overall M\&C system to make this interface as common as possible across the ngEHT sites.

As the number of stations and observations grows, providing on-site VLBI expertise will become increasingly challenging. The ngEHT design approach follows an operations model where station operators can remotely perform any required operations and maintenance, with specialist support being provided only when necessary. Remote operation is facilitated by the focus on human factors and user-centered design, and leads to less reliance on manual operations and analysis. A cloud-based deployment of the array-level M\&C system is envisioned as the way to provide "operations from anywhere" capability to the array operations staff. This is expected to include server, database, and UI components that facilitate operation of the array. M\&C capability at each station is still required to provide the control inputs to station subsystems and aggregate the local data for relay to the array-level system. Remote access to both the array- and station-level M\&C systems are provided with appropriate security, authentication, and authorization methods.

To achieve all this, the M\&C system architecture is expected to be built from off-the-shelf software components using open standards, including databases, message queueing and information exchange methods, and user interface frameworks. This facilitates development and maintenance over the lifecycle of the array. A robustly defined software architecture allows isolation of site-specific dependencies to the smallest and fewest components necessary.

\subsection{Antennas} \label{subsec:Antennas}

\begin{figure}[t]
    \centering
    \includegraphics[angle=270,width=0.9\columnwidth]{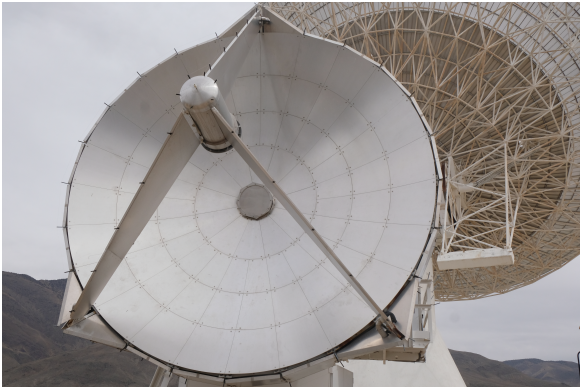}
    \caption{Photograph showing the condition of the BIMA antenna dish surface (from March 2022)}
    \label{fig:BIMAdishPanels}
\end{figure}

The ngEHT concept adds $\sim$10 new antennas to the existing EHT array. In \phaseone the ngEHT program will deploy 3-4 modest diameter antennas for the most rapid increase in next-generation science (see Section~\ref{sec:SiteSelection}). To mitigate risk, the program has identified two possible paths towards this \phaseone enhancement.  The first would use three 6~m diameter antennas from the decommissioned Berkeley-Illinois-Maryland-Array (BIMA), which would be transported to the Las Campanas, San Pedro Martir, and Canary Island sites. 

The BIMA dishes have a surface accuracy specification of $\sim$40 $\mu$m rms, sufficient for operation up to 345 GHz. Photogrammetry measurements will allow re-adjustment the surface to the required accuracy after re-assembly of the antenna. The panels of all three dishes are in good condition, as shown in Figure \ref{fig:BIMAdishPanels}.

\begin{table} \centering
\begin{tabular}{ll}
\hline
\multicolumn{2}{c}{Design specifications} \\
\hline
Primary reflector diameter & 9 m \\
Mount architecture & Alt-Az \\
Optics & Cassegrain \\
Sun avoidance zone & None \\
\hline
\multicolumn{2}{c}{Operating specifications}\\
\hline
Surface accuracy & 30 $\mu$m rms\\
Frequency range & 86 --- 345 GHz\\
Aperture efficiency & 0.8\\
Pointing accuracy & 2$''$ rms (all sky, blind)\\
Tracking accuracy & $0.''2$\\
Aperture blockage & $<5\%$\\
Gain variation with elevation & $<5\%$\\
Range of motion in azimuth & -180$^{\circ}$ --- 360$^{\circ}$ \\
Range of motion in elevation & 3$^{\circ}$ --- 90$^{\circ}$\\
Slew speed & 1$^{\circ}$/s\\
\hline
\multicolumn{2}{c}{Environmental specifications}\\
\hline
Temperature & -15C---35C operational\\
            & -20C---45C high\\
            & -30C---55C survival\\
Wind speed & 10 m/s operational\\
           & 15 m/s high\\
           & 50 m/s survival\\
\hline
\end{tabular}
\caption{Specifications of the new ngEHT antennas}
\label{table:antennaSpecs}
\end{table}

Figure \ref{fig:anchor_plot} suggests that a 6~m diameter antenna with an aperture efficiency of 0.8 would allow us to reach the required sensitivity, when paired with a large collecting area dish such as LMT or phased ALMA. But a larger diameter antenna will relax the requirement on long distance baselines away from such anchor stations, and also have two additional advantages: easier calibration for pointing and focus measurements, and ability to carry out single-dish science projects while the antennas are not observing for ngEHT in VLBI mode.

Therefore, a second possible \phaseone implementation path would be to use newly fabricated dishes of 9~m diameter.  The specifications of the new antennas are summarized in Table \ref{table:antennaSpecs}.  The ngEHT team is in discussions with several telescope vendors and it is clear that dishes with the required specifications can be procured within a reasonable cost envelope.  In this case, \phaseone would target four sites: the Mt. Jelm site in Wyoming, in addition to Canary Islands, San Pedro Martir and Las Campanas.

\section{Summary and conclusions} \label{sec:Summary}
The ngEHT, described initially to the Astro2020 decadal survey review \citep{Doeleman_2019}, is a program to plan extensions of the EHT array that will deliver high dynamic range imaging and movie making capability for black hole studies on event horizon scales.  It does so principally by deploying modest-diameter radio dishes at optimized geographical locations, which significantly increases interferometric baseline coverage (Figures~\ref{fig:eht2022_globe_uv},~\ref{fig:ngeht_phase1_globe_uv},~\ref{fig:ngeht_phase2_globe_uv}), by implementing a simultaneous tri-band (86, 230, 345~GHz) receiver suite, and increasing the bandwidth of backend systems and data processing pipelines.

The process and initial results of optimizing site selection for ngEHT telescopes described here indicates two possible paths to achieve a next-generation EHT array.

In the first path, \phaseone consists of adding dishes at two existing sites (OVRO and Haystack) to the current EHT, and available refurbished dishes from the BIMA array would be relocated to three sites (Las Campanas, Chile; San Pedro Martir, Mexico; Canary Islands, Spain).  Then in \phasetwo, additional sites would be developed; current analysis indicates that the combination of these locations: La Paz, Bolivia; Wyoming, US; Marangu, Tanzania; Santiago, Chile; and Bern, Switzerland, constitute an array that can deliver all of the threshold Key Science Goals.  These \phasetwo sites should be considered possibilities at this stage; more work is required to assess them at all levels, including thorough consideration of cultural and environmental aspects.  

In an alternate path, \phaseone would again add both OVRO and Haystack to the EHT, and four new 9~m diameter dishes would be deployed to the Mt. Jelm site in Wyoming; Las Campanas, Chile; San Pedro Martir, Mexico; and Canary Islands, Spain.  Then in \phasetwo,  planned new telescopes are added to the array as they become available, including the AMT, LLAMA and KVNYS, KVNPC facilities. 
\textit{Either of these approaches to realizing the ngEHT leads to the increases in global array capabilities that are required to achieve all ngEHT Key Science Goals.}

Strategies for ngEHT data transport, correlation, calibration and data reduction are all developed.  Requirements for major instrumental sub-systems are specified, and details of prototypes to be used are described.  In sum, this work brings the ngEHT project to the point of readiness for implementation.

\acknowledgments

Support for this work was provided by the NSF through grants AST-1952099, AST-1935980, AST-1828513, and AST-1440254, and by the Gordon and Betty Moore Foundation through grant GBMF-10423.  This work has been supported in part by the Black Hole Initiative at Harvard University, which is funded by grants from the John Templeton Foundation and the Gordon and Betty Moore Foundation to Harvard University.  

\bibliography{references}{}
\bibliographystyle{aasjournal}

\appendix
\numberwithin{equation}{section}

\setcounter{table}{0}
\renewcommand{\thetable}{A\arabic{table}}

\section{Additional site selection details} \label{app:SiteDetails}

\autoref{tab:CandidateSites} lists the sites considered for the ngEHT array optimization procedures described in \autoref{sec:OptimizingArray}.  This pool of candidate sites has been taken from \cite{Raymond_2021}, and they have been selected for their favorable atmospheric transmission properties at 230\,GHz and 345\,GHz during the typical EHT observing season in March and April.

\begin{table*}
    \begin{tabular}{lcccc}
    \hline
    Site code & Location & Latitude & Longitude & Elevation (m) \\
    \hline
    ALI	&	Hotan County, China	&	35.963	&	79.338	&	6080 \\
    BAN	&	Alberta, Canada	&	51.350	&	-116.206	&	3470 \\
    BAR	&	California, US	&	37.634	&	-118.256	&	4340 \\
    BGA	&	Progled, Bulgaria	&	41.695	&	24.738	&	1730 \\
    BGK	&	Westfjords, Iceland	&	66.032	&	-23.052	&	830 \\
    BLDR	&	Colorado, US	&	39.588	&	-105.643	&	4340 \\
    BMAC	&	Eastern Cape, South Africa	&	-31.096	&	27.889	&	2420 \\
    BOL	&	La Paz, Bolivia	&	-16.351	&	-68.131	&	5230 \\
    BRZ	&	Esp\'irito Santo, Brazil	&	-20.439	&	-41.799	&	2850 \\
    CAS	&	Tierra del Fuego, Argentina	&	-54.790	&	-68.415	&	2850 \\
    CAT	&	R\'io Negro, Argentina	&	-41.170	&	-71.486	&	2100 \\
    CNI &   La Palma, Canary Islands    &   28.299 &   -16.509 &   2360 \\
    DomeA	&	Upper ice sheet, Antarctica	&	-80.367	&	77.351	&	4090 \\
    DomeC	&	Upper ice sheet, Antarctica	&	-75.101	&	123.342	&	3230 \\
    DomeF	&	Upper ice sheet, Antarctica	&	-77.317	&	39.702	&	3700 \\
    ERB	&	Khalifan, Iraq	&	36.584	&	44.466	&	2110 \\
    FAIR	&	Alaska, US	&	64.988	&	-147.599	&	620 \\
    FLWO	&	Arizona, US	&	31.675	&	-110.951	&	1270 \\
    FUJI	&	Fujinomiya \& Yamanashi, Japan	&	35.367	&	138.730	&	3750 \\
    GARS	&	Trinity Peninsula, Antarctica	&	-63.320	&	-57.895	&	20 \\
    GLTS	&	Ice sheet summit, Greenland	&	72.580	&	-38.449	&	3230 \\
    HAN	&	Ladakh, India	&	32.780	&	78.963	&	4500 \\
    JELM	&	Wyoming, US	&	41.097	&	-105.977	&	2940 \\
    KEN	&	Meru, Kenya	&	-0.141	&	37.315	&	4260 \\
    KILI	&	Kilimanjaro, Tanzania	&	-3.088	&	37.406	&	4430 \\
    LCO &   Coquimbo, Chile &   -29.032  &   -70.685 &   2320 \\
    LOS	&	New Mexico, US	&	35.880	&	-106.675	&	2000 \\
    NOB	&	Nagano, Japan	&	35.944	&	138.472	&	1370 \\
    NZ	&	Canterbury, New Zealand	&	-43.987	&	170.465	&	1010 \\
    ORG	&	Oregon, US	&	42.635	&	-118.576	&	2970 \\
    PAR	&	Antofagasta, Chile	&	-24.628	&	-70.404	&	2640 \\
    PIKE	&	Colorado, US	&	38.841	&	-105.041	&	4280 \\
    SAN	&	California, US	&	34.099	&	-116.825	&	3500 \\
    SGO	&	Santiago, Chile	&	-33.3346	&	-70.270	&	3350 \\
    SKS	&	Crete, Greece	&	35.212	&	24.898	&	1740 \\
    SPM    &   Baja California, Mexico &   31.045  &   -115.464    &   2800 \\
    SPX	&	Fieschertal, Switzerland	&	46.548	&	7.985	&	3510 \\
    SUF	&	Zaamin, Uzbekistan	&	39.623	&	68.468	&	2440 \\
    TRL	&	Jutulsessen, Antarctica	&	-72.010	&	2.540	&	1280 \\
    VLA &   New Mexico, US  &   34.079  &   -107.618 &   2120 \\
    YAN	&	Huanca Sancos, Peru	&	-13.938	&	-74.392	&	4230 \\
    YBG	&	Lhasa Tibet, China	&	30.006	&	91.027	&	5360 \\
    \hline
    \end{tabular}
    \caption{List of candidate sites for the ngEHT, updated from \cite{Raymond_2021}.}
    \label{tab:CandidateSites}
\end{table*}

\autoref{tab:BaseArrays} specifies the pre-existing arrays assumed during the site selection procedure described in \autoref{sec:OptimizingArray}.  Four different variants of pre-existing array are explored as parameters in the site selection procedure, and these variants are enumerated in the table.

\begin{table*}
    \begin{tabular}{lp{6cm}c}
    \hline
    Parameter set & Pre-existing stations from EHT array & Other pre-existing stations assumed \\
    \hline
    \phaseone set 1 & none & HAY, OVRO \\
    \phaseone set 2 & LMT & HAY, OVRO \\
    \phaseone set 3 & APEX, GLT, JCMT, LMT, SMT & HAY, OVRO \\
    \phaseone set 4 & ALMA, APEX, GLT, IRAM, JCMT, KP, LMT, NOEMA, SMA, SMT, SPT & HAY, OVRO \\
    \hline
    \phasetwo set 1 & none & CNI, HAY, LCO, OVRO, SPM \\
    \phasetwo set 2 & LMT & CNI, HAY, LCO, OVRO, SPM \\
    \phasetwo set 3 & APEX, GLT, JCMT, LMT, SMT & CNI, HAY, LCO, OVRO, SPM \\
    \phasetwo set 4 & ALMA, APEX, GLT, IRAM, JCMT, KP, LMT, NOEMA, SMA, SMT, SPT & CNI, HAY, LCO, OVRO, SPM \\
    \hline
    \end{tabular}
    \caption{The different pre-existing arrays considered as part of the site selection exploration (\autoref{sec:SiteSelection}).  Each of these combinations of stations is the starting set of sites for which the addition of three sites (for the \phaseone analysis) or five sites (for the \phasetwo analysis) are explored.  These starting arrays are chosen to generally represent the possible operating modes shown in \autoref{tab:OpsModes}.  Set 1, for example, might be a minimal array useful for Target of Opportunity observations.  Sets 2 and 3, with the addition of a large aperture, could provide flexible long-term monitoring capability.  And set 4 includes all possible stations for a full campaign mode.  The range of starting arrays also give some indication of optimal placement in the full campaign mode in the case where some sites are not available due to weather or technical issues.}
    \label{tab:BaseArrays}
\end{table*}

\end{document}